\documentclass[preprint2]{aastex}

\slugcomment{Accepted by ApJ in September 2011}

\shorttitle{Spitzer-IRAC Emission in Herbig-Haro Objects (II)}
\shortauthors{Takami et al.}

\begin{document}



\title{A Detailed Study of Spitzer-IRAC Emission in Herbig-Haro Objects (II): Interaction Between Ejecta and Ambient Gas}


\author{Michihiro Takami\altaffilmark{1}, Jennifer L. Karr\altaffilmark{1}, Brunella Nisini\altaffilmark{2}, Thomas P. Ray\altaffilmark{3}
}

\altaffiltext{1}{Institute of Astronomy and Astrophysics, Academia Sinica.
P.O. Box 23-141, Taipei 10617, Taiwan, R.O.C.; hiro@asiaa.sinica.edu.tw}
\altaffiltext{2}{INAF --- Osservatorio Astronomico di Roma Via di Frascati 33, 00040, Monteporzio Catone, Italy}
\altaffiltext{3}{Dublin Institute for Advanced Studies, 31 Fitzwilliam Place, Dublin 2, Ireland}


\begin{abstract}
We present a new analysis of the physical conditions in three Herbig-Haro complexes (HH 54, HH 212, and the L 1157 protostellar jet) using archival data from
the Infrared Array Camera (IRAC) on the {\it Spitzer Space Telescope}. As described in detail in Paper I, the emission observed using the 4.5-\micron~filter is enhanced in molecular shocks ($T$=1000--4000 K) at relatively high temperature or densities compared with that observed with the 8.0-\micron~filter. Using these data sets, we investigate different distributions of gas between high and low temperatures/densities. Our analysis reveals the presence of a number of warm/dense knots, most of which appear to be associated with working surfaces such as the head of bow shocks and cometary features, and reverse shocks in the ejecta. 
These are distributed not only along the jet axis, as expected, but also across it.  While some knotty or fragmenting structures can be explained by instabilities in shocked flows, others can be more simply explained by the scenario that the mass ejection source acts as a ``shot gun'', periodically ejecting bullets of material along similar but not identical trajectories. 
Such an explanation challenges to some degree the present paradigm for jet flows associated with low-mass protostars.
It also give clues to reconciling our understanding of the mass ejection mechanism in high and low mass protostars and evolved stars. 
\end{abstract}





\section{Introduction}

Collimated jets have been observed in a number of low-mass protostars over a variety of stellar masses and evolutionary stages. These jets have traditionally been seen via the thermal excitation of atoms and ions in partially ionized shocked regions, i.e., line emission at optical wavelengths \citep[H$\alpha$, [O{\scshape i]} etc., see][for a review]{Bally07}. More recently, jets associated with the youngest protostars have been observed in molecular emission at radio (e.g., CO, SiO, SO) and infrared wavelengths (e.g., H$_2$) \citep[see][for reviews]{Arce07, Bally07}.

The popular magneto-centrifugal models beautifully explain the driving and collimation of these jets \citep[][]{Shu00,Konigl00}. Magnetic fields are assumed to be coupled
either to the star and star-disk interface (in the so-called X-wind model) or the disk surface (in the disk wind theory).
At the inner edge or surface of the accretion disk, the gas is lifted from the disk and accelerated due to a combination of centrifugal force and magnetic fields aligned with the disk. Such a flow will spin, and at larger distances the spinning motion will twist the magnetic field, providing a ``hoop stress" which collimates the outflowing gas.  Possible non-steady alternative mechanisms for the driving include magnetic pressure \citep[e.g., ][]{Uchida85,Machida08} and stress \citep[e.g.,][]{Hayashi96, Goodson99}. Although the driving mechanism is not yet clear, astronomers are reaching the consensus that jets are magneto-hydrodynamically driven and powered by mass accretion.

In their simplest form some of the above theories predict that  the ejecta of the jet are distributed as a continuous flow or, as seems more likely due to unsteady mass accretion, a chain of knotty structures moving in opposite directions.  In reality, the observations of protostellar jets show a much more complex variety of features (knots, bow shocks, bubbles etc.) than predicted by idealized models. These features are usually attributed to hydrodynamic/magnetohydrodynamic instabilities of the flows, or time variability of the ejection in mass, velocity and/or direction \citep[][]{Bally07,Arce07}.
Mass ejection that is variable in mass and velocity can be attributed to instabilities in the accretion disk \citep[e.g.,][]{Bell94,Balbus98} or the episodic reconnection of magnetic field lines between the protostar and disk \citep[e.g.,][]{Hayashi96,Goodson99}.

While low-mass protostars tend to exhibit collimated jets, high-mass protostars tend to show outflows with wide opening angles \citep[][]{Arce07}. Studies of warm shocked gas show that such outflows, particularly the most energetic ones, are highly turbulent. The best known example is the OMC-1 outflow, which consists of many chains of knots (ÒfingerÒ structures) which have been ejected into the ambient cloud \citep[][]{Allen93,Nissen07}. In contrast, the Cepheus A outflow complex is associated with chains of bow shocks in near-infrared H$_2$ emission, and these can be explained with a pulsed, precessing jet \citep{Cunningham09}. It is likely that outflows associated with high-mass protostars are also powered by mass accretion, however, it is not clear how their driving mechanism is related to that of the collimated jets seen toward low-mass protostars \citep[][]{Arce07}.

In general,
observational studies of the driving of protostellar jets and outflows have been hampered by  several issues. Their driving engines are expected to be too compact  to be directly resolved without interferometers \citep[i.e., they are within 2-3 AU of the protostar, corresponding to 15--20 milliarcsec in the nearest star forming regions at 140 pc ---][]{Shu00,Konigl00,Machida08,Hayashi96,Goodson99}. Moreover the ``engine'' is often heavily embedded, limiting the observing wavelengths to radio and far infrared. Finally, in many cases we are not directly observing the ejected gas, but rather the gas that is shocked or compressed due to interactions between the jet and ambient gas \citep[][]{Bally07,Arce07}.


The InfraRed Array Camera (IRAC) on the Spitzer Space Telescope has been used to study a variety of protostellar jets and outflows \citep[e.g.,][]{Noriega-Crespo04,SmithH06,Walawender06,Davis07,Qiu08,Ybarra09,Neufeld09}. \citet[][hereafter Paper I]{Takami10b} investigated the nature of infrared emission with IRAC in six Herbig-Haro complexes. In Paper I we showed that thermal H$_2$ emission at $T$=1000--4000 K can explain the morphological similarities and differences observed between the four bands, and also the spectral-energy distributions (SEDs) observed in some regions. SEDs in other regions show excess emission at 4.5 \micron, presumably because of excitation of the CO band at high densities. See also \citet[][]{Smith05b,Ybarra09,Neufeld08} for similar studies. The observations and LTE/non-LTE calculations of thermal H$_2$ emission indicate that the emission at 3.6 and 4.5 \micron~tends to be associated with regions at high temperatures or densities, while that at 8.0 \micron~tends to be associated with regions at low temperatures or densities. 
Paper I also suggests that the $I_{4.5 \micron}/I_{3.6 \micron}$ ratio is useful for discriminating between Herbig-Haro knots and foreground/background stars: this ratio is, $\ga$1.6 for Herbig-Haro knots; $\sim$0.6 for foreground stars with $A_K$=0; $\sim$1.1 for background stars with $A_K$=5. 

In this paper we extend our analysis for three objects (the L 1157 protostellar jet, HH 54, and HH 212) to understand their interaction with the ambient gas and investigate the nature of their mass ejection. We analyze the distribution of shocked gas at high temperatures/densities using the differing spatial distributions of emission at 4.5 and 8.0 \micron. The remaining part of the paper is organized as follows. In \S 2 we summarize the data set we have used. In \S 3 we describe our approach for discriminating between warm/dense and cool/diffuse regions, and show how these are distributed in the individual objects. In \S 4 we discuss their physical conditions based on the IRAC colors, and also the spectra obtained using the Spitzer InfRared Spectrograph (IRS) for the L 1157 jet \citep{Neufeld09,Nisini10}. In \S 5 we discuss the origin of these structures, and implications for the distribution of the ejecta from the source and their driving mechanism. In \S6 we present our conclusions.

\section{Data}

Archival data in four IRAC bands (3.6, 4.5, 5.8, and 8.0 \micron) were obtained for the L 1157 jet \citep[$d \sim$250 pc,][]{Looney07}, HH 54 \citep[$d \sim$200 pc, see][]{caratti06}, and HH 212 \citep[$d \sim$400 pc, see][and references therein]{LeeC08}. All the data had been reduced with the post-BCD pipelines developed by IPAC. The mean FWHMs of the point response functions (PRFs) are 1.66, 1.72, 1.88, and 1.98'' for the four bands, respectively.
Previous publications with these data include \citet{Looney07,Neufeld09} for the L 1157 jet, and \citet{Ybarra09} for HH 54. These data sets were also used in Paper I to investigate the nature of shocked emission observed in IRAC bands. The median of the total integration time per pixel ranges from 21--322 s depending on the target.

\section{Image Analysis}

\subsection{Method}
Based on Paper I, we expect the $I_{4.5 \micron}/I_{8.0 \micron}$ ratio allows us to discriminate regions at high temperature or densities from the others since the 4.5-\micron~emission is more enhanced than the 8.0-\micron~emission. In other words,  a high $I_{4.5 \micron}/I_{8.0 \micron}$ ratio would be observed in regions at high temperature or density. However, this flux ratio can suffer severely from diffuse extended PAH emission at 8.0 \micron~\citep[e.g.,][]{Reach06,Looney07}. 
When flux ratio maps are calculated, masking the low signal to noise regions is standard practice, due to the relatively large uncertainties yielding very large (positive and negative) ratios. For our purposes, this means that we could not easily compare the regions of shocked emission to the background regions in order to determine the origin of the flux ratio. 
%
%

Therefore, we alternatively apply the following approach: subtracting a flux-scaled 8.0-\micron~image from a 4.5-\micron~image ($I_{4.5 \micron} - r \times I_{8.0 \micron}$, where $I_{4.5 \micron}$ and $I_{8.0 \micron}$ are intensity at 4.5 and 8.0 \micron, respectively, and $r$ is the scaling factor). The scaling factor $r$ corresponds to the critical $I_{4.5 \micron}/I_{8.0 \micron}$ ratio for discriminating between regions with high and low $I_{4.5 \micron}/I_{8.0 \micron}$: the regions with $I_{4.5 \micron}/I_{8.0 \micron} > r$ show $I_{4.5 \micron} - r I_{8.0 \micron}$ values larger than the background; those with $I_{4.5 \micron}/I_{8.0 \micron} < r$ show $I_{4.5 \micron} - r I_{8.0 \micron}$ values smaller than the background. The advantage of this method over the $I_{4.5 \micron}/I_{8.0 \micron}$ map is that it is applicable even for regions with non-uniform diffuse background emission, in particular for HH 212 among our objects; and some regions in the L 1157 jet. This is because (1) such maps allow us to display the background regions without S/N masks; and (2) the measurements can be made relative to the value for the adjacent background. This is in principle the same approach as that used for decades to investigate ionization and inferred shock conditions in 
Herbig-Haro objects observed in 
atomic and ionic lines~\citep{Reipurth92,Heathcote96,Heathcote98,Fridlund98,Hartigan00,Reipurth02,McCoey04}.

Figure \ref{fig1} shows an example of analysis in the HH 54 bow-shock region, which does not clearly show the presence of diffuse PAH emission in any band. 
Before obtaining the $I_{4.5 \micron}/I_{8.0 \micron}$ and $I_{4.5 \micron} - r I_{8.0 \micron}$ maps, the median value of the background is subtracted from each image, and the 4.5- and 8.0-\micron~images are convolved with the PRF at 8.0 and 4.5 \micron, respectively. The PRFs at these two wavelengths are significantly different due to different diffraction patterns and degree of scattered light in the detector, and this approach is in particular useful for accurately comparing their fluxes at the same angular resolution (FWHM$\sim$3" after convolution). Figure \ref{fig1} shows that the $I_{4.5 \micron} - r I_{8.0 \micron}$ images allows us to successfully identify regions of high $I_{4.5 \micron} /I_{8.0 \micron}$ ratios. Applying different $r$'s (i.e., different critical $I_{4.5 \micron}/I_{8.0 \micron}$ ratios to show positive and negative values) merely changes the contrast for displaying the features in shocks, as shown in later sections for the other regions as well.

In the $I_{4.5 \micron} - r I_{8.0 \micron}$ images the foreground and background stars tend to show positive values as do shocked regions at high temperatures or densities. We discriminate between these two kinds of features using (1) conventional three color images (blue, green and red for 3.6, 4.5 and 8.0 \micron, respectively), and (2) $I_{3.6 \micron}/I_{4.5 \micron}$ flux ratio maps. The stars tend to appear blue in the three-color image, and exhibit a large $I_{3.6 \micron}/I_{4.5 \micron}$ flux ratio (Paper I). The example for the HH 54 bow shock region is shown in Figure \ref{fig1}. In Appendix A we identify all the foreground and background stars in our regions of interest.

\subsection{Results}

Figure \ref{fig2} shows our whole view of the L 1157 jet, HH 212, and HH 54. Figure \ref{fig3} shows close-up views and the $I_{4.5 \micron} - r I_{8.0 \micron}$ images of the sub-regions selected from L 1157 and HH 212 in Figure \ref{fig2}. As for HH 54 in Figure \ref{fig1}, the median value of the background is subtracted from each image, and the 4.5- and 8.0-\micron~images are convolved with the PRF at 8.0 and 4.5 \micron, respectively, before obtaining the  $I_{4.5 \micron} - r I_{8.0 \micron}$ maps. We mark prominent features in shocks in which  $I_{4.5 \micron} - r I_{8.0 \micron}$ is higher than the adjacent background using a letter and a number. 
 For each region, the scaling factor $r$ (i.e., the critical $I_{4.5 \micron}/I_{8.0 \micron}$ value) is arbitrarily adjusted for contrast between the emission regions with high densities/temperatures and the remaining regions. The different $r$'s required in different regions results from varying shock conditions, yielding different average or median $I_{4.5 \micron} /I_{8.0 \micron}$ ratios. This is independent of the diffuse background, as quantitatively proven in \S 4.

The $I_{4.5 \micron} - r I_{8.0 \micron}$ images, in particular for L 1157 A, show a number of faint and unmarked point-like features with $I_{4.5 \micron} - r I_{8.0 \micron}$ higher than the adjacent background. These are likely due to faint stars whose signal-to-noise for $I_{4.5 \micron}$ is less than 10 (see Appendix). In the text we limit our discussion to  features clearly identified as shocks with this criteria for signal-to-noise.
Some regions in Figure \ref{fig3} show global variation of $I_{4.5 \micron} - r I_{8.0 \micron}$ in the background region due to the diffuse PAH emission apparent in the $I_{8.0 \micron}$ band. 

The $I_{4.5 \micron} - r I_{8.0 \micron}$ images in Figures \ref{fig1}  and \ref{fig3} exhibit a number of knotty features in shocks with $I_{4.5 \micron} - r I_{8.0 \micron}$ higher than the background. As discussed in \S 2, these are the regions with relatively high temperature or density.
The peak positions of the knots are listed in Table \ref{tbl1}.
%
%
%
%
As described below in detail, these are categorized into: (1) those bracketed by bow shocks or bubble-like structures with small $I_{4.5 \micron} - r I_{8.0 \micron}$ values (i.e., emission at relatively low temperatures or densities) ; (2) those located at the head of bow shocks;  and (3) the others. Most are associated with bright knotty structures in the three-color image.

In the HH 54 A, L 1157 A and C regions, HH 54 A1--A4  and L 1157 A1/A2/C1 are bracketed by bow shocks or bubble-like structures with small $I_{4.5 \micron} - r I_{8.0 \micron}$ values, i.e., emission at relatively low temperatures or densities. In HH 54 A, four prominent knots are identified in two bow shocks which overlap. 
HH 54 A1 and A3 are located at the head of two bow shocks. The positions of HH 54 A1--A4 match peaks in the 4.5-\micron~image identified by \citet{Ybarra09} (I, B, E, A, respectively, in their Figure 3.)
In the L 1157 A region, multiple knotty structures are observed in A1 and A3, and each of A1 and A3 is bracketed by an asymmetric bow-shock-like structure at low temperature or density. In the L 1157 C region a single knot with a high temperature or density (C1) is bracketed by a single bubble-like emission feature at low temperature or density.
Among the warm/dense knots described above, HH 54 A2,  A4,  L 1157 A3 and  C1 exhibit corresponding bright yellow knotty structures in the three-color images. 

The three-color image of the the HH 212 A and B regions exhibits  bow shocks with angular scales of $\sim$30", and the warm/dense knots 
 A1 and B1 are respectively associated with their heads. Another two warm/dense knots (A2, B2) are observed in  the west wake of the HH 212 A/B bow shocks, respectively. The remaining part of the wakes, including the east wake of the HH 212 A/B bow shocks, show $I_{4.5 \micron} - r I_{8.0 \micron}$ values smaller than the background, indicating that these regions have relatively low $I_{4.5 \micron}/I_{8.0 \micron}$  and thereby low temperatures or densities. The morphology in the upper stream of A2 and B2 suggest they are the heads of asymmetric small bow shocks \citep[see also the H$_2$ 1-0 S(1) high resolution images obtained by, e.g., ][]{Zinnecker98, Takami06b}.
 
Although Figure \ref{fig3} is useful for identifying the remaining warm/dense knots (L 1157 A3/A4/B1--B5/C2 and HH 212 C1--C3), it does not clearly show how these are associated with gas of lower temperatures or densities. Figure \ref{fig4} shows a close-up view of L 1157 A3, B1, B3, B5, C2, and HH 212 C3.
This figure shows that these warm/dense knots are associated with tails of smaller $I_{4.5 \micron} - r I_{8.0 \micron}$, thereby of lower temperature or density, towards the upper stream.

Figure \ref{fig5} shows the distribution of warm/dense knots over the entire region of the L 1157 jet and HH 212. The distribution of warm/dense knots is asymmetric between the northern and southern lobes of the L 1157 jet, as expected from previous observations of this object \citep[e.g.,][]{Bachiller01,Looney07}. A number of knots are observed in the northern lobe, distributed within an opening angle of about 30$^\circ$. The regions with relatively low temperatures or densities, which look bright in the $I_{4.5 \micron} - r I_{8.0 \micron}$ image, show a relatively continuous morphology analogous to the CO $J$=2--1 emission observed by \citet{Bachiller01}. In contrast, only two bright warm/dense knots are identified in the southern lobe. The bubble-like morphology is not clearly observed in the CO $J$=2--1 image by \citet{Bachiller01}, but is apparent in the  CO $J$=1--0 image by \citet{Gueth96} at an angular resolution similar to our study.


In contrast to the L 1157 jet, the warm/dense knots in HH 212 show a simpler distribution in Figure \ref{fig5}.  A1, B1, C1, C3 are distributed within 1$^\circ$ of a specific P.A. (24/204$^\circ$), indicating a straight jet axis. In addition to the known positional symmetry between A1 and B1 \citep[NK3 and SK3 in][]{Zinnecker98}, A2 and B2 show an intriguing symmetry with respect to the jet axis. Both are located at similar distances from the protostar (72" and 65", respectively) and with similar offset angles from the jet axis (7$^\circ$), but A2 is offset clockwise while B2 counterclockwise.

\section{Spectral Energy Distributions and Comparisons with IRS Spectra}
To investigate the different physical conditions between warm/dense knots and cool/diffuse regions, we measured the fluxes in the four IRAC bands at the positions shown in Figures \ref{fig6}--\ref{fig8}. These include all the warm/dense knots we identified in Figures \ref{fig1} and \ref{fig3} except HH 212-C1, in which the shocked emission is severely contaminated by non-uniform background emission at 5.8 and 8.0 \micron. For each position, we carefully subtracted the diffuse background emission by measuring it in the $x$- and $y$-directions and fitting it using a linear function. We then measured the average flux density in a circular 6" aperture.
Table \ref{tbl2} shows the measured fluxes at these positions. The uncertainty shown in Table \ref{tbl2}  are due to non-uniform background emission in addition to the shot noise and readout noise of the detector.  According to the IRAC Data Handbook 3.0, the absolute flux of extended emission measured using this camera is highly uncertain, and this depends on the spatial distribution of the emission both in and outside the aperture. This fact is not included in the uncertainties listed in Table \ref{tbl2}. One would expect those additional uncertainties to be $\sim 20$ \% for the 3.6- and 4.5-\micron~fluxes; $\sim 30$ \% for 5.8-\micron; $\sim 40$ \% for 8.0-\micron; $\sim 0.3$ mag. for the [4.5]-[8.0] color; and $\sim 0.2$ mag. for [5.8]-[8.0], based on the flux correction factor for different apertures listed in the IRAC Data Handbook.

Figure \ref{fig9} shows the spectral energy distributions (SEDs) at each position marked in the individual regions (L 1157 A/B/C, HH 54, HH 212 A/B/C). The SEDs for warm/dense positions show shallower slopes than those in cool/diffuse positions in most of the regions. Some or all of the SEDs for the warm/dense positions in L 1157 B, L 1157 C, HH 54, HH 212 A and HH 212 B show excess emission at 4.5 \micron, while the other SEDs are relatively straight in Figure \ref{fig9}. 
In HH 212 C the differences in SED between warm/dense and cool/diffuse regions are marginal.

These SEDs are similar to those measured in the six jet systems (including the L 1157 jet, HH 54, and HH 212) in Paper I. In Paper I we showed that (1) the SEDs without 4.5-\micron~excess can be explained well by shocked H$_2$ emission with a power law cooling function ($\Lambda \propto T^{-\alpha}$); (2) the different SED slopes can be attributed to different power indexes or densities; and (3) those with 4.5-\micron~excess require a contaminant at this wavelength, presumably CO, indicative of higher densities than those without 4.5-\micron~excess. To investigate this in detail, we plot the observed and modeled colors in color-color diagrams in Figures \ref{fig10} and \ref{fig11}. As in Paper I, models for H$_2$ emission are made for the following cases: (1) iso-thermal cases, and shock slabs with a power-law cooling function ($\Lambda \propto T^{ - \alpha}$); (2) thermal collisions with H+He and H$_2$+He, corresponding to the cases with relatively high ($\gg 0.002-0.02$) and low dissociation rates ($\ll 0.002-0.02$, see Paper I). As modeled in Paper I, we adopt  $A$-coeffcients provided by \citet{Wolniewicz98}, and collisional rate coefficients for H$_2$ and He by \citet{LeBourlot99}. For collisional rate coefficients with H, we adopt \citet{Wrathmall07} and \citet{LeBourlot99}, and show the results separately. According to \citet{Wrathmall07}, they provide rate coefficients with a better accuracy than \citet{LeBourlot99} due to improved representation of the vibration eigenfunctions. In contrast, coefficients provided by \citet{LeBourlot99} can explain the observed IRAC colors better in Paper I. The fluxes for  H$_2$+He collisions are calculated with the lowest 36 energy levels (32 lines in the four IRAC bands), while those for  H+He collisions are done with the lowest 49 levels (45 lines in the four IRAC bands). The number of transitions we include is limited by the availability of collisional rate coefficients. The level populations of ortho- and para-H$_2$ are calculated separately, and the resultant fluxes are combined assuming an ortho/para ratio of 3. The applicability and limitations of those calculations are discussed in Paper I in detail.

Figures \ref{fig10} and \ref{fig11} show that, as in Paper I, the models with a power-law cooling function explain the observed colors at the majority of the above positions. In Figure \ref{fig11}, they are located near the colors for local thermal equilibrium (LTE) with a power index of the cooling function $\alpha$ of 3--6. The colors measured at most of the remaining positions show [4.8]--[8.0]  smaller than the models, which is presumably due to contaminating emission from CO in the 4.5-\micron~band (Paper I). As shown by \citet{Neufeld08}, the contribution of CO to the entire 4.5-\micron~flux is significant at densities $\ga 10^7$ cm$^{-3}$. The [4.5]--[8.0] colors larger than the LTE positions, indicative of lower densities, are observed at a few cool/diffuse positions in HH 54. Figure \ref{fig10} shows that the colors measured at all the positions have [4.5]--[8.0] smaller than the isothermal cases by at least 1--2 mag.

In Figure \ref{fig11}, the [5.8]--[8.0] and [4.5]--[8.0] colors measured in each position show a correlation similar to the models with different power law indexes $\alpha$. In particular, the warm/dense positions show smaller [5.8]--[8.0] and [4.5]--[8.0] colors, corresponding to small $\alpha$. This can be explained if the warm/dense knots include a larger fraction of warm gas that cool/diffuse regions shown in the $I_{4.5\micron}/I_{8.0\micron}$ images. In the HH 54 region, the colors at the warm/dense positions indicate that the gas density is sufficiently high for LTE, while those at the cool/diffuse positions indicate lower densities. In Figures \ref{fig10} and \ref{fig11}, we also show the $I_{4.5\micron}/I_{8.0\micron}$ ratios corresponding to the scaling factor (critical flux ratio) $r$ in Figures \ref{fig1} and \ref{fig3}. In each region, the [4.5]--[8.0] color corresponding to the middle $r$ is larger than those at warm/dense positions, and smaller than those at cool/diffuse positions. This implies that the $I_{4.5\micron}/I_{8.0\micron}$ ratios at the warm/dense positions are larger than the middle $r$, while those at cool/diffuse regions are lower. All the  results described above are consistent with our argument in \S 3.1 that  $I_{4.5 \micron} - r I_{8.0 \micron}$ 
allows us to discriminate between these warm/dense and cool/diffuse regions, independent of diffuse background at 8.0 \micron.

\citet{Neufeld09} and \citet{Nisini10} have made Spitzer IRS observations of the L 1157 jet at the 5.2--37 \micron~range, and analyze the physical conditions using H$_2$ S(0) to S(7) lines. In Table \ref{tbl3} we compare our results in the same outflow with theirs in the following manner: 
(1) comparing the LTE temperature inferred by the H$_2$ 0--0 S(5)/S(6)/S(7) lines \citep{Nisini10} with that obtained from the observed $I_{5.8\micron}/I_{8.0\micron}$ ratio; 
(2) estimating the IRAC 8.0-\micron~flux assuming that the flux is dominated by H$_2$ 0--0 S(4) and S(5) lines \citep{Neufeld08,Neufeld09} and comparing with the observed IRAC flux at the same position; and 
(3) estimating the IRAC 4.5-\micron~flux using the 0-0 S(9)/S(5) flux ratio, assuming that the 0-0 S(9) is responsible for all the IRAC 4.5-\micron~flux, and in LTE with the temperature provided by \citet{Nisini10}. The positions for the comparisons listed in Table \ref{tbl3} are selected based on signal-to-noise for the IRS spectra for the measurement of temperature in Figure 5 of \citet{Nisini10}.

In Table \ref{tbl3} the temperatures obtained from the IRS spectra and IRAC colors agree at most of the positions. In each region both temperatures are higher for the warn/dense knots than those for cool/diffuse regions we identified in Section 3.
Furthermore, the IRAC 8.0-\micron~fluxes estimated from the 0--0 S(4) and S(5) lines approximately agree with the observed fluxes listed in Table \ref{tbl3}.
These support our discussion regarding modeled colors which assumes that the emission at 5.8- and 8.0-\micron~is dominated by H$_2$, in particularly the 0--0 S(4) to S(7) lines \citep[cf., ][]{Smith05b,Neufeld08, Neufeld09,Ybarra09}.
In contrast, the IRAC 4.5-\micron~fluxes estimated in the above manner are systematically lower than the observed fluxes, by a factor of 0.06--0.3. This discrepancy can be attributed to a combination of the following for the estimated flux: (1) assumption of iso-temperature, which yields systematically large [4.5]--[8.0] colors (thereby low $I_{4.5 \micron}/I_{8.0 \micron}$ flux ratios) than shock models with a power-law cooling function (Figures \ref{fig10} and \ref{fig11}); (2) the fact that the other H$_2$ lines and CO emission are not included.

\section{Discussion}
As described in \S 3 in detail, most of the warm/dense knots shown in Figures \ref{fig1}, \ref{fig3}, and \ref{fig4} are categorized into: (1) those bracketed by bow shocks or bubble-like structures seen in regions at relatively low temperatures or densities ; (2) those located at the head of bow shocks;  and (3) those associated with cometary tails. Throughout \S 5.1, we suggest that these are located at working surfaces where the ejecta from the protostar interact with the ambient gas. In \S 5.2 we discuss the implications for the distribution of the ejecta from the protostar.

\subsection{Interaction between the jet and ambient gas}
As shown in \S 3, several of the warm/dense knots we identified (HH 54 A1--A4, L 1157 A1, A2, C1) are bracketed by a bow shock or bubble-like feature at lower temperature or density. A likely explanation for these warm/dense knots is that these are reverse shocks into the ejecta from the protostar \citep[i.e., Mach disks or cloudlet shocks --- see, e.g., ][ for schematic views]{Hartigan89}. Such shocks have been observed towards a few protostellar jets seen in atomic or ionic emission \citep[see][for a review]{Reipurth01}.

\citet[][]{Reipurth92,Fridlund98, Reipurth02} have shown different excitations between Mach disks and forward bow shocks using the H$\alpha$ and [S{\scshape ii}] lines. In HH 34, H$\alpha$ is more enhanced in the bow shock while [S{\scshape ii}] is more enhanced in the Mach disks, and this can be explained if the ejecta from the gas has a higher density than the ambient gas, yielding lower ionization \citep[][]{Reipurth92,Reipurth02}. In contrast, the opposite trend is observed in the L 1551 bow shock \citep[][]{Fridlund98}. The higher temperatures or densities in HH 54 A1--A4, L 1157 A1, A2, and C1 coupled with the bow shock or bubble-like feature can be attributed to the different pre-shock densities of the ejecta from the protostar and the ambient gas.


In HH 212 A and B the emission at high temperatures or densities is associated with the heads of bow shocks, while that at low temperature or density is associated with its wake (Figure \ref{fig3}). This agrees with previous studies of bow shocks. Since we expect higher shock velocities at the head, this gas should have higher temperatures (and perhaps higher densities) at these positions \citep[][for a review]{Hartigan87, Reipurth01}. The structures seen in HH 212 A and B suggest the presence of a relatively small bow shock (A2, B2) in the wake of a larger bow shock (A1, B1). The structure of the small bow shocks is not clear at the angular resolution of our study, but is seen more clearly in the H$_2$ 2.12 \micron~images at higher angular resolutions \citep[][]{Zinnecker98,Takami06b}.

None of the bow shocks associated with HH 212 A1/A2/B1/B2 clearly show evidence for the presence of reverse shocks, in contrast to HH 54 A1--A4, L 1157 A1, A2, and C1. This may be attributed to a number of reasons. First, the reverse shocks can be significantly fainter if their preshock density is significantly lower than that of the ambient gas \citep[by a factor of more than 10--100 --- see][]{Hartigan89}. Second, reverse shocks may be relatively unstable, and thereby highly variable, as observed in HH 47 in optical emission lines \citep{Hartigan05}.

Cometary structures like those associated with L 1157 A3/B1/B3/B5, L 1157 C2, and HH 212 C3 are rarely seen in protostellar jets. Indeed, none of the reviews of protostellar jets over the past decade has reported them \citep[][]{Hartigan00, Eisloffel00, Reipurth01,Bally07, Ray07}.
Even so, it is natural to interpret these as ejecta interacting with the ambient gas at these positions, providing a higher temperature or density and thereby enhancing the 4.5-\micron~emission. The tails, which have a temperature or density lower than the head, could be gas entrained due to this interaction.

\subsection{Implications for jet driving}

According to the present jet paradigm, the ejecta of the jet is distributed as either a continuous flow or a chain of knotty structures, moving in opposite directions \citep[e.g.][]{Shu00,Konigl00,Hayashi96,Goodson99}. As already stated, the observations of protostellar jets show a much more complex variety of features (knots, bow shocks, bubbles etc.) than the idealized models. These features are often attributed to hydrodynamic/magnetohydrodynamic instabilities of the shocked flows \citep[e.g.,][]{Vishniac83,Wardle90,Stone95},  or time variability of the ejection in mass and direction \citep[e.g.,][]{Raga04_HH32,Raga04_HH4647,Raga07b,Raga09}.

Some warm/dense knots show distributions similar to the fragmented structures seen in hydrodynamical simulations.
The Kelvin-Helmholz and/or Rayleigh-Talor instabilities can fragment the ejecta along and across the flow axis  \citep[e.g.,][]{Ouyed03, Klein94}, and these could explain fragmented structures in each of HH 54 A, L 1157 A1--A3 and HH 212 C3. Perhaps cometary structures in the L 1157 jet could be attributed to fragmentation similar to that seen in wind \citep[e.g.,][]{Cunningham05}, bullet \citep[e.g.,][]{Raga07a}, or  collimated pulsed jet with precession \citep[e.g.,][]{Rosen04b,Smith07}.
In contrast, none of these simulations show the following observed features. These are:- (1) two bow shocks nearly overlapping with each other in HH 54; (2) bubble-like and cometary features parallel to each other in the L 1157 C region; (3) symmetry between HH 212 A2 and B2 about the jet axis. Furthermore, it is not clear why the same possible instability in the L 1157 B region does not break up the flow structures further downstream (i.e., the L 1157 A region).

The symmetry between HH 212 A2 and B2 about the jet axis is particularly notable, since the high degree of spatial symmetry along the jet axis of HH 212 also appear to rule out the possibility that the features are imposed on an originally uniform jet via flow instabilities \citep[][]{Zinnecker98}. \citet[][]{Zinnecker98} pointed out that it is much more likely that those features arise through time variability at the driving source. Similarly, the symmetry between HH 212 A2 and B2 can be easily explained if it is due to the driving source.

The above features in HH 54 A, L 1157 C and HH 212 A/B would therefore require an alternative explanation. One possibility is that  the mass ejection from these protostars is better approximated by ballistic bullets ejected in roughly (but not exactly) opposite directions. Indeed, such a scenario can easily explain the observed structures described above.
Although the detailed mechanism for such mass ejection is not clear, it has been proposed for the high mass protostellar outflow OMC-1 \citep[e.g.,][]{Allen93,Nissen07} and proto-planetary nebulae \citep[e.g.,][]{Dennis08}. In the case of low-mass protostellar jets, \citet{Hartigan05} revealed a number of knotty structures in the well-studied bow shock HH 47 A, and argue that these are explained as small bullets that pass from the jet through the Mach disk and working surface to emerge as bumps in the bow shock. \citet{Yirak09} have recently questioned the stability of a continuous magnetized fluid, and alternatively simulates the jet using a model similar to the one we propose above. 

If the trajectories of the bullets are approximately the same as the arrows shown in Figure \ref{fig4}, the faintness of some features (L 1157 A3 and B5) can be explained as follows. The ambient gas is swept up by shocks associated with A1 and A2. The passage of these shocks then allows the ambient gas to move downstream, thereby weakening the impact of A3 and B5. Throughout, these features are consistent with the scenario of the shot-gun-like mass ejection discussed above.

\section{Conclusions}
We have analyzed data for three Herbig-Haro complexes (the L 1157 jet, HH 212, HH 54) obtained using the Infrared Array Camera (IRAC) on the {\it Spitzer Space Telescope} to investigate how these flows interact with the ambient gas. A combination of 3.6-, 4.5- and 8.0-\micron~images has allowed us to successfully identify a number of warm/dense knots in shocked emission in molecular gas at 1000--4000 K. Most of them are categorized into: (1) those bracketed by bow shocks or bubble-like structures seen in regions at relatively low temperatures or densities; (2) those located at the head of bow shocks;  and (3) those associated with cometary tails. Our results show that these locations are distributed not only along the jet axis, as expected, but also across it.  

It is likely that the first group of warm/dense knots are associated with the ejecta from the protostar (i.e., Mach disks or cloudlet shocks). Analogous to some optically-visible Herbig-Haro objects, different $I_{4.5 \micron}/I_{8.0 \micron}$ ratios between the reverse and forward shocks can be attributed to the different conditions of the pre-shock gas, i.e., the ejecta from the protostar and  ambient gas. The third group is explained if the ejecta interacts with the ambient gas at the location of the warm/dense knots, providing a relatively high shock velocity and thus compressing the gas. Throughout, we suggest that all of the above features are located at the working surfaces where the ejecta from the protostar interact with the ambient gas. 

The origin of some knotty structures can be attributed to shock instabilities previously modeled by other authors. In contrast, others do not seem to be easily attributed to the same physical processes with a continuous flow or a chain of well aligned knotty structures. A simple alternative explanation for these observations, in particular for the HH 54 A, L 1157 C and HH 212 A-B regions, is that the mass ejection is more akin to a ``shot gun'', periodically ejecting bullets of material along similar but not identical trajectories.
This explanation challenges to some degree the present paradigm for protostellar jet flows as quasi-continuous,
but give clues to reconciling our understanding of the mass ejection mechanism in high and low mass protostars and evolved stars. 

\acknowledgments

We are grateful to Dr. D. Neufeld for use of their IRS spectra, and  to Dr. F. Shu, R. Krasnopolsky, C.-F. Lee, O. Morata, M. Cemeljic, N. Hirano, and S. Takakuwa for useful discussions.
We also thank the anonymous referee for useful comments.
The IRAC images were obtained through the Spitzer Archive operated by IPAC.
This research made use of the SIMBAD data base operated at CDS, Strasbourg, France, and  dNASA's Astrophysics Data System Abstract Service. This study is supported from National Science Council of Taiwan (Grant No. 97WIA0100327).



{\it Facilities:} \facility{Spitzer Space Telescope (IRAC)}.



\appendix
\section{Identification of foreground and background stars}
\begin{figure*}
\epsscale{2}
\plotone{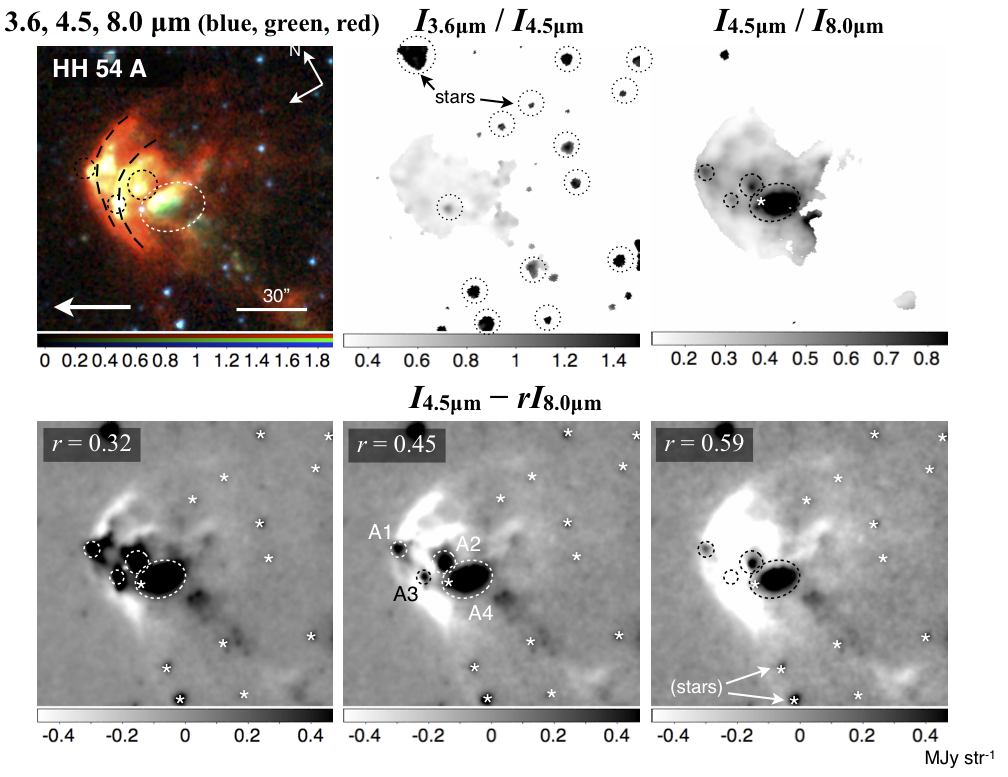}
\caption{
The HH 54 bow shock region.
({\it top-left}) three-color image (blue, green, red for 3.6, 4.5 and 8.0 \micron, respectively)
({\it top-middle}) the $I_{3.6 \micron}/I_{4.5 \micron}$ flux ratio map. The ratio is shown in regions where the signal-to-noise is larger than 10 for $I_{4.5 \micron}$.
({\it top-right}) $I_{4.5 \micron}/I_{8.0 \micron}$ 
({\it bottom}) $I_{4.5 \micron} - r I_{8.0 \micron}$ images with different $r$'s.
The warm or dense knots in shocks (i.e. those with large $I_{4.5 \micron} - r I_{8.0 \micron}$ and
$I_{4.5 \micron}/I_{8.0 \micron}$ values) prominent in the $I_{4.5 \micron} - r I_{8.0 \micron}$ and
$I_{4.5 \micron}/I_{8.0 \micron}$ images are marked with dashed circles and ellipses. Stars identified based on the $I_{3.6 \micron}/I_{4.5 \micron}$ ratio and the three-color image (see \S 2 for details) are marked with dotted circles in the  $I_{3.6 \micron}/I_{4.5 \micron}$ map, and white asterisks in the $I_{4.5 \micron} - r I_{8.0 \micron}$ maps. Dashed curves in the three-color image indicate two bow shocks discussed in
Sections 3 and 4.
\label{fig1}}
\end{figure*}

Figure \ref{figA1} shows the foreground and background stars identified in the L 1157 A, B, and C regions. As described in the caption of Figure \ref{fig1}, identification of these stars is based on their point-source appearance and at least one of the following: (1) a blue color in the three-color image, and (2) a higher $I_{3.6 \micron}/I_{4.5 \micron}$ flux ratio than the surrounding region. The last is based on the detailed study of shocked emission in Herbig-Haro objects of Paper I ($\la$0.6 for Herbig-Haro knots; $\sim$1.6 for foreground stars with $A_K$=0; $\sim$0.9 for background stars with $A_K$=5\footnote{This column density is higher than the maximum column density measured in some molecular clouds using background stars \citep{Chapman09}. See also \citet{Motte01} for the molecular envelopes of low-mass  protostars.}), and here we apply this to the regions where the signal-to-noise is larger than 10 for $I_{4.5 \micron}$. Contaminating emission from shocks can yield a lower $I_{3.6 \micron}/I_{4.5 \micron}$ flux ratio for some stars, hence we apply a more qualitative identification with the other criteria.

For HH 212, the 3.6-\micron~image suffers significantly from non-uniform diffuse extended emission, which would affect identification of foreground and background stars using the $I_{3.6 \micron}/I_{4.5 \micron}$ flux ratio map. This diffuse emission component is also seen at 8.0 \micron~with a very similar distribution, and can be removed reasonably well if we scale the flux of the 8.0-\micron~image and subtract it from the 3.6-\micron~image.

Figure \ref{figA2} shows how this process works: i.e., the image at 3.6 \micron, that at 8.0 \micron~the flux of which is scaled to a level similar to 3.6 \micron, and the 3.6-\micron~image from which the scaled 8.0-\micron~image is subtracted.
Foreground and background stars are faint in the scaled 8.0-\micron~image as expected: they should be 2\% and 4\% of the 3.6-\micron~ flux for most spectral types (i.e., [3.6]--[8.0]$\sim$0) and extinction $A_K$=0 and 5, respectively, based on \citet{Chapman09}. In contrast, shocked emission is apparent at 8.0 \micron. This implies our imaging process keeps the stellar fluxes high, while it degrades shocked emission. Throughout, this would allow an even clearer discrimination between shocks and foreground/background stars using the $I_{3.6 \micron}$/$I_{4.5 \micron}$ flux ratio (the exact formula for this case is $(I_{3.6 \micron}-a \times I_{8.0 \micron}$)/$I_{4.5 \micron}$, where $a$ is the scaling factor for the 8.0-\micron~flux). Figure \ref{figA3} shows the foreground and background stars identified in the HH 212 A, B, and C regions, replacing the $I_{3.6 \micron}$/$I_{4.5 \micron}$ images with $(I_{3.6 \micron}-a \times I_{8.0 \micron}$)/$I_{4.5 \micron}$ images. The scaling factor $a$ was determined using the least squares method to minimize the residual of subtraction for the non-uniform diffuse extended emission.

The most likely origin of the non-uniform diffuse extended emission is PAHs illuminated at the surface of the molecular cloud. The measured scaling factor $a$ of 0.071 (HH 212 A) and 0.0821 (HH 212 B,C), which correspond to the $I_{3.6 \micron}$/$I_{8.0 \micron}$, is in excellent agreement with recent models for UV-excited PAH emission by \citet[][]{DL07} (0.063--0.084 for $U$=1--10 and 0.059--0.083 for $U$=100, where $U$ is the UV field normalized to the local interstellar radiation field).

All the stars identified in Figure \ref{fig1} are seen in the 2MASS images for HH 54, for which the images with a long exposure (6x) are available. Most of the stars in Figure \ref{figA1} and some of the stars in Figure \ref{figA3} are seen in the 2MASS images, while the remaining faint stars are not, due to the significantly shallower detection limit of 2MASS.





\clearpage



\begin{figure*}
\epsscale{2}
\plotone{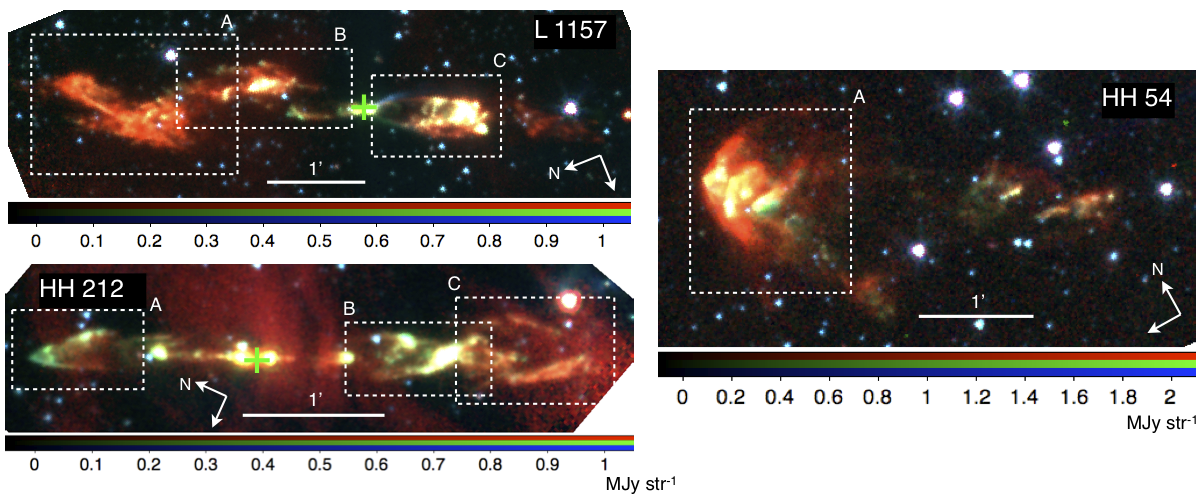}
\caption{
Three-color images (blue, green, red for 3.6, 4.5 and 8.0 \micron, respectively)  for the L 1157 jet, HH 54 and HH 212.
Dashed boxes show the regions where we apply the analysis described in \S 2.
Green crosses in the L 1157 and HH 212 regions indicate the position of the protostar, based on \citet{Bachiller01} and \citet{LeeC08}, respectively. The driving source of HH 54 has not been clearly identified \citep{caratti06}
\label{fig2}}
\end{figure*}

\clearpage

\begin{figure*}
\epsscale{2.1}
\plotone{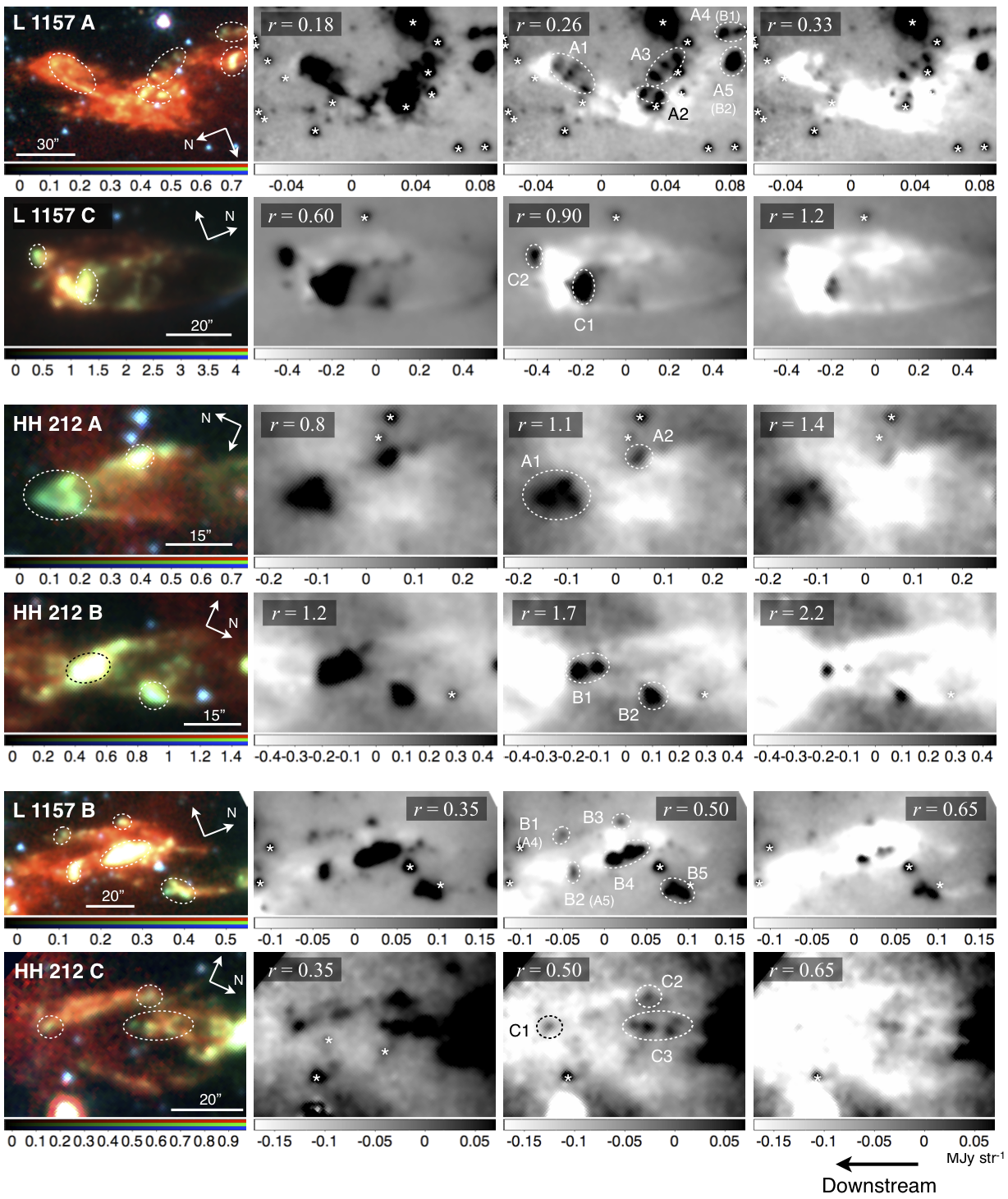}
\end{figure*}

\clearpage

\figcaption{
Three-color images (blue, green, red for 3.6, 4.5 and 8.0 \micron, respectively) and
$I_{4.5 \micron} - r I_{8.0 \micron}$ maps with different $r$ for the regions selected in Figure \ref{fig2} (for L 1157, HH 212).
All images are oriented to make left the downstream direction.
Dashed ellipses show the warm/dense knots in shocks prominent in the $I_{4.5 \micron} - r I_{8.0 \micron}$ maps. The corresponding regions are also marked in the three-color image. These are labeled with a combination of a letter corresponding to
the region (A--C), and a number (1--5).  L 1157 A4 and A5 are identical with B1 and B2, respectively, but are shown with
different $r$ values for the maps of L 1157 A and B.
The asterisks show stars identified using the three-color image and $I_{3.6 \micron}/I_{4.5 \micron}$ flux ratio
(see \S 2 and Appendix).
\label{fig3}}

\clearpage

\begin{figure*}
\epsscale{2.0}
\vspace{-1cm}
\plotone{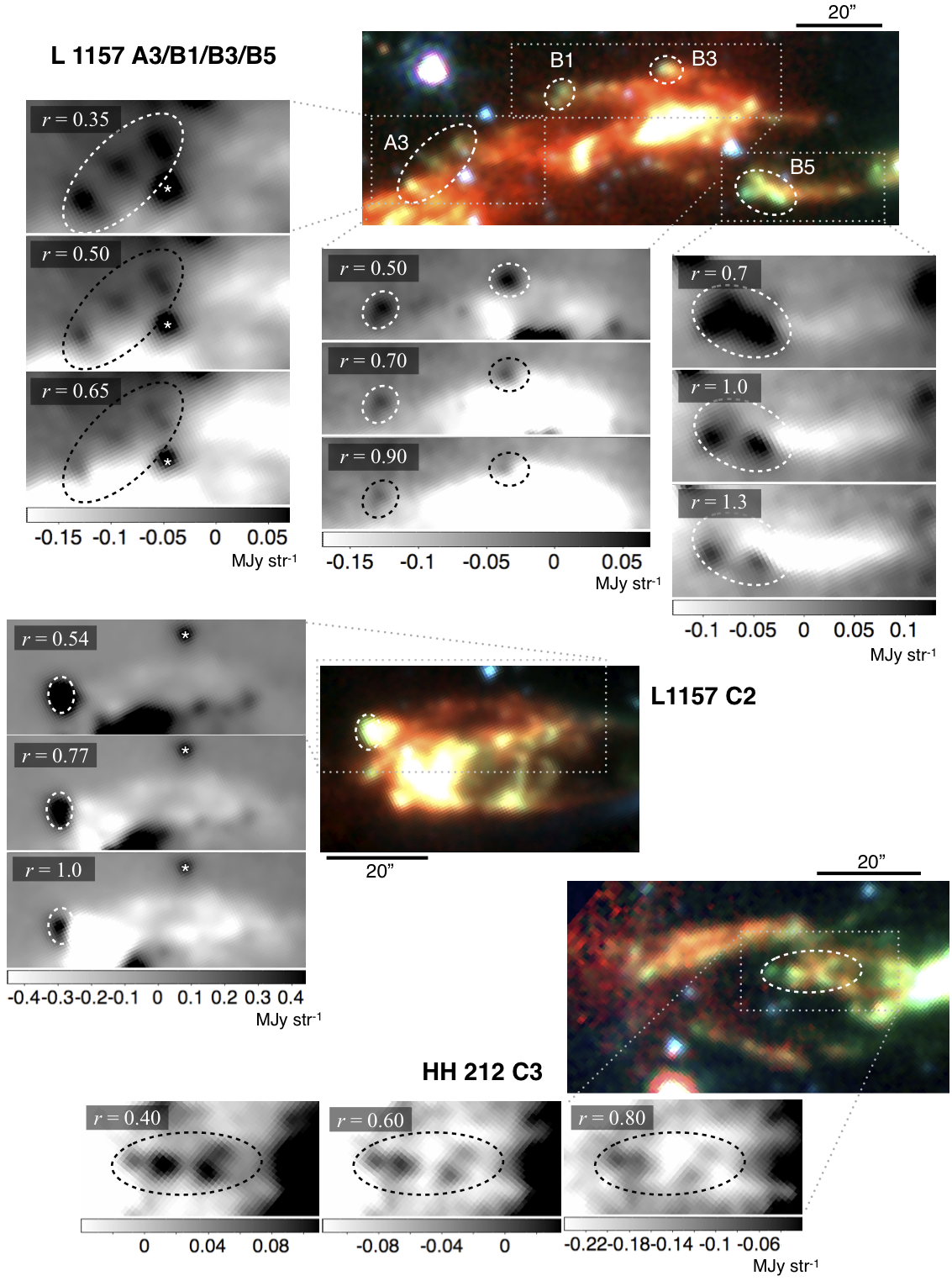}
\end{figure*}

\clearpage

\figcaption{
Same as Figure \ref{fig3} but for some warm/dense knots associated with tails (L 1157 A3/B1/B3/B5/C2, HH 212 C3). For most of the regions the value of $r$ has been changed to show the contrast between the head and tail(s) of cometary structures.
\label{fig4}}

\clearpage

\begin{figure*}
\epsscale{2.0}
\vspace{-1.3cm}
\plotone{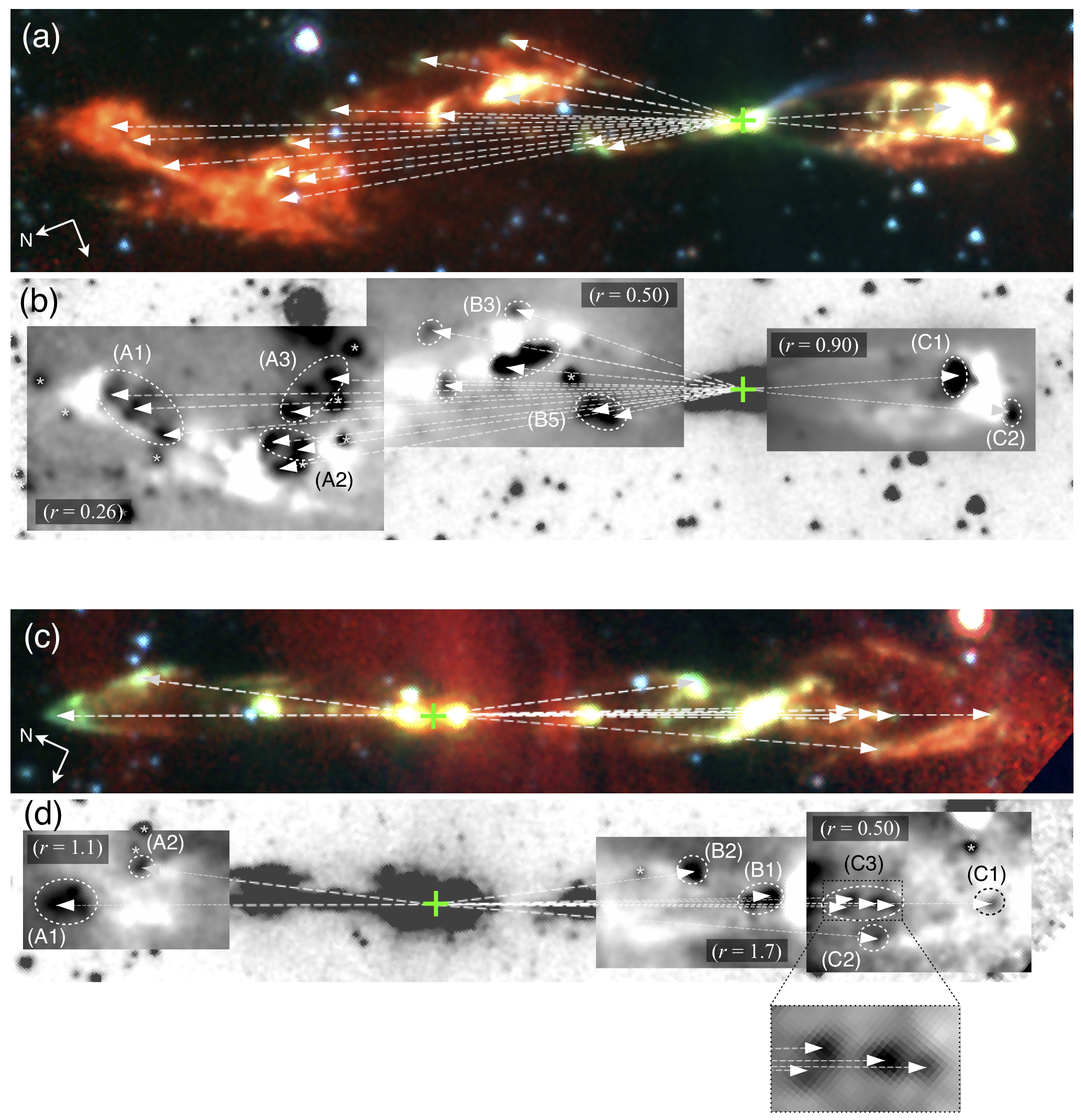}
\vspace{-1cm}
\caption{
(a) Three-color image of L 1157. The arrows indicate the position of warm/dense knots marked in Figure \ref{fig3} relative to the position of the protostar \citep{Bachiller01} marked with the green cross. (b) The $I_{4.5 \micron} - r I_{8.0 \micron}$ images in Figure \ref{fig3} are superimposed on the 4.5-\micron~image in gray scale of the same region. The warm/dense knots are labeled in the same manner as Figure \ref{fig3}, and the arrows are the same as (a). The asterisks show the stars. (c)(d) Same as (a)(b) but for HH 212. The position of the protostar is based on \citet{LeeC08}.
\label{fig5}}
\end{figure*}

\clearpage

\begin{figure*}
\epsscale{2.0}
\plotone{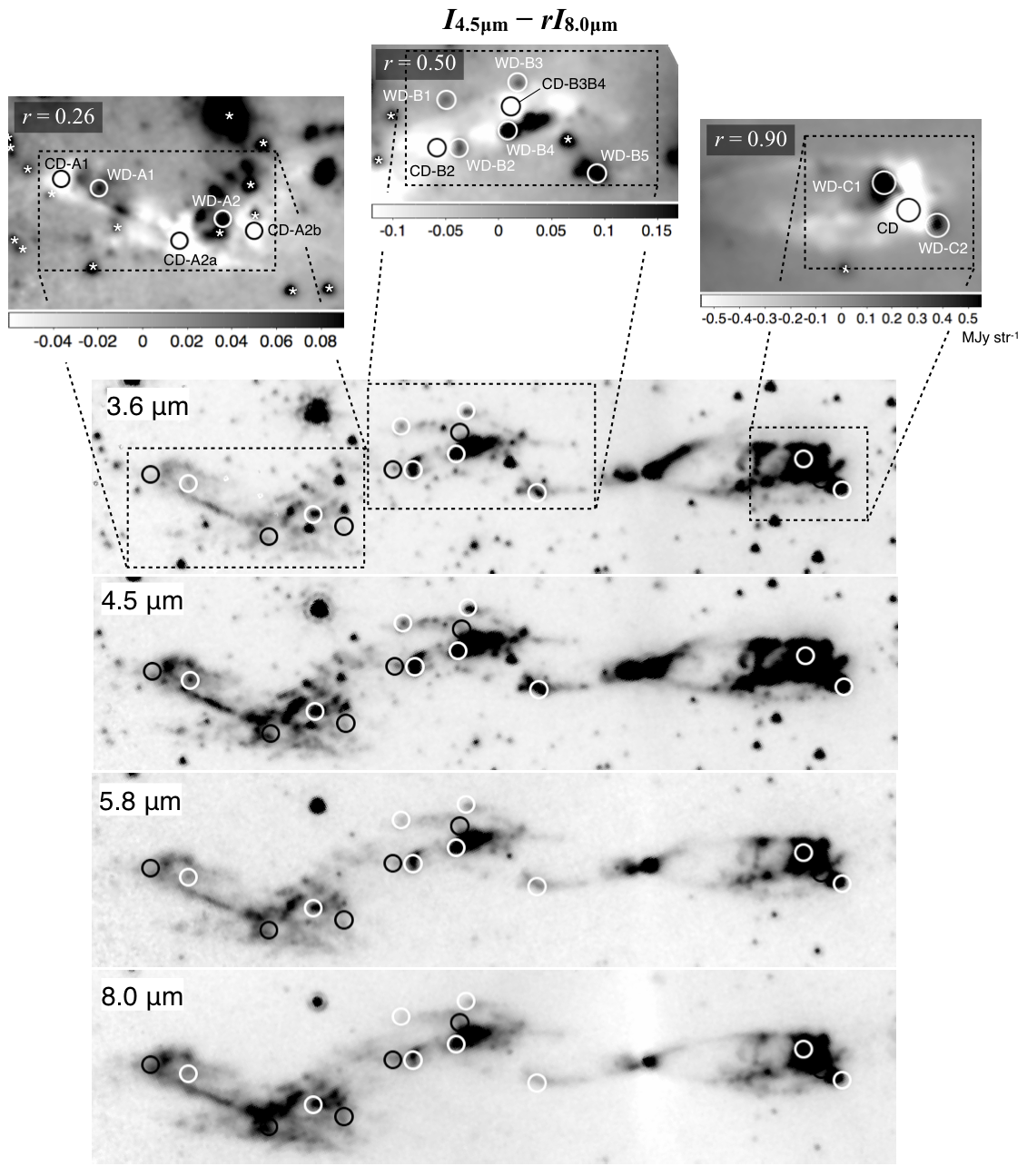}
\end{figure*}

\clearpage

\figcaption{Positions for circular aperture photometry (6") in the L 1157 jet, on the $I_{4.5 \micron} - r I_{8.0 \micron}$ and 3.6-, 4.5-, 5.8-, and 8.0-\micron~images. 
The white and black circles are the apertures for warm/dense and cool/diffuse components, respectively. These are labeled as follows: WD --- warm/dense position shown in the $I_{4.5 \micron} - r I_{8.0 \micron}$ image; CD --- cool/diffuse position shown in the $I_{4.5 \micron} - r I_{8.0 \micron}$ image; A1/A2/B1-B5/C1/C2 --- positions (or adjacent cool/diffuse positions) for the warm/dense knots identified in Figure \ref{fig3}.
Dashed boxes and lines show positional coincidence between the $I_{4.5 \micron} - r I_{8.0 \micron}$ and 3.6-\micron~images. Stars also show values larger than the surrounding region in the $I_{4.5 \micron} - r I_{8.0 \micron}$, and these are marked in these images using asterisks (see \S 2 and Appendix A for identification).
\label{fig6}}

\clearpage

\begin{figure*}
\epsscale{2.0}
\plotone{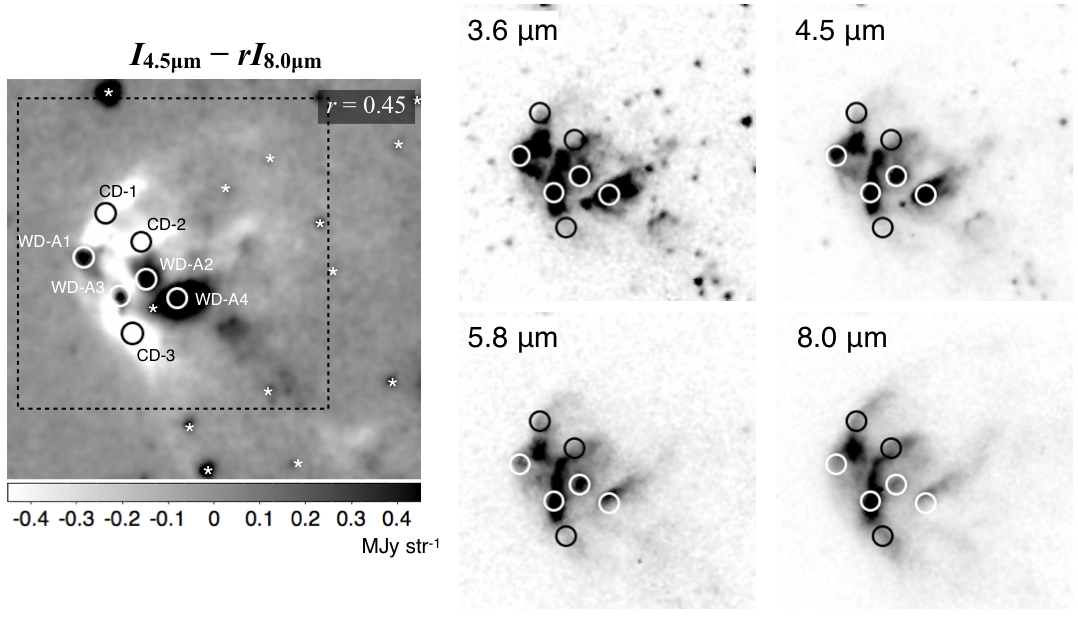}
\caption{Same as Figures \ref{fig6} but for HH 54. A1--A4 in the labels are based on identification of warm/dense knots in Figure \ref{fig1}. See Appendix A for identification of stars in the $I_{4.5 \micron} - r I_{8.0 \micron}$ image.
\label{fig7}}
\end{figure*}

\clearpage

\begin{figure*}
\epsscale{2.0}
\plotone{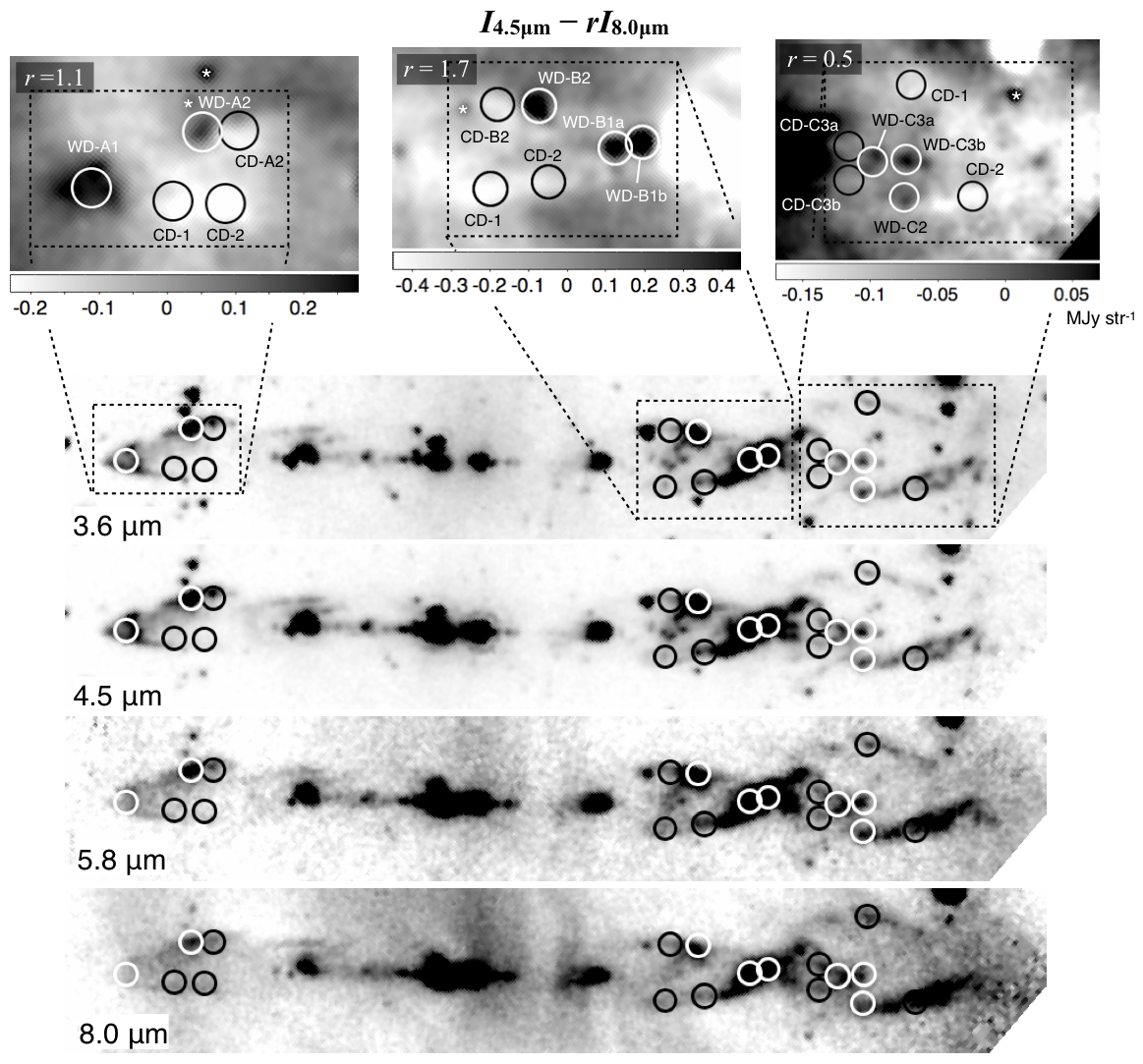}
\caption{Same as Figure \ref{fig6} but for HH 212.
\label{fig8}}
\end{figure*}

\clearpage

\begin{figure*}
\epsscale{2.0}
\plotone{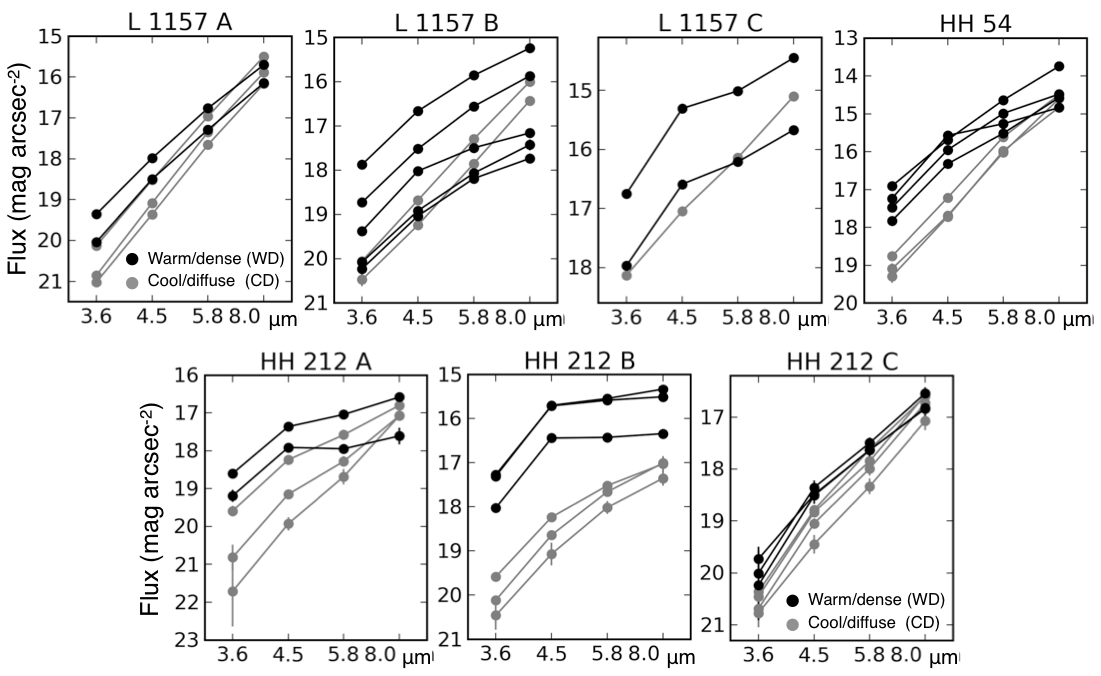}
\caption{Spectral energy distributions for the four IRAC bands at the positions shown in Figures \ref{fig6}--\ref{fig8}. Solid and gray lines are those measured in the warm/dense knots (WD) and cool/diffuse regions (CD), respectively. Error bars are shown only for those larger than the size of the dots. These are based on Table \ref{tbl2}, and possible systematic errors for the absolute flux calibration (see text) are not included.
\label{fig9}}
\end{figure*}

\clearpage

\begin{figure*}
\epsscale{2.15}
\vspace{-3.2cm}
\plotone{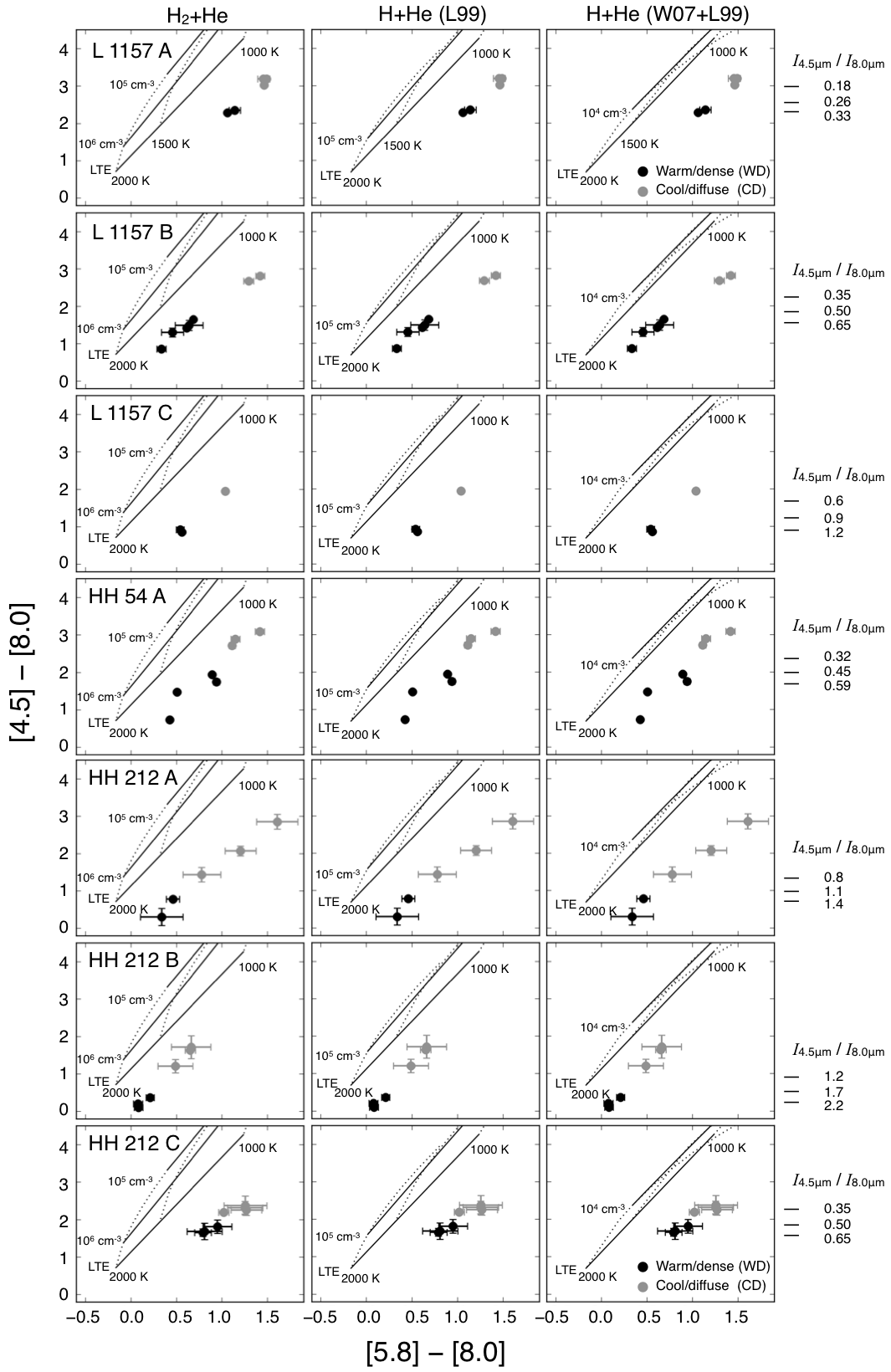}
\vspace{-0.5cm}
\end{figure*}

\clearpage

\figcaption{
Color-color diagram for observed fluxes in the individual regions, and isothermal H$_2$ emission. Solid and gray dots indicate the colors measured in warm/dense and cool/diffuse regions, respectively. Error bars are shown only for those larger than the size of the dots. These are based on Table \ref{tbl2}, and possible systematic errors for the absolute flux calibration (see text) are not included. Models are shown for the following three cases from left to right: ($left$) thermal collisions with H$_2$ and He, adopting the collisional rate coefficients provided by Le Boutlot et al. (1999);  ($middle$) same but collisions with H and He;  ($right$) same as the middle but adopting the rate coefficients provided by Wrathmall et al. (2007) for collisions with H. The [4.5]--[8.0] colors corresponding to $r$ (i.e., the critical $I_{4.5\micron}/I_{8.0\micron}$ ratios) in Figures \ref{fig1} and \ref{fig3} are shown at the right side of each row.
\label{fig10}}

\clearpage

\begin{figure*}
\epsscale{2.15}
\vspace{-3.2cm}
\plotone{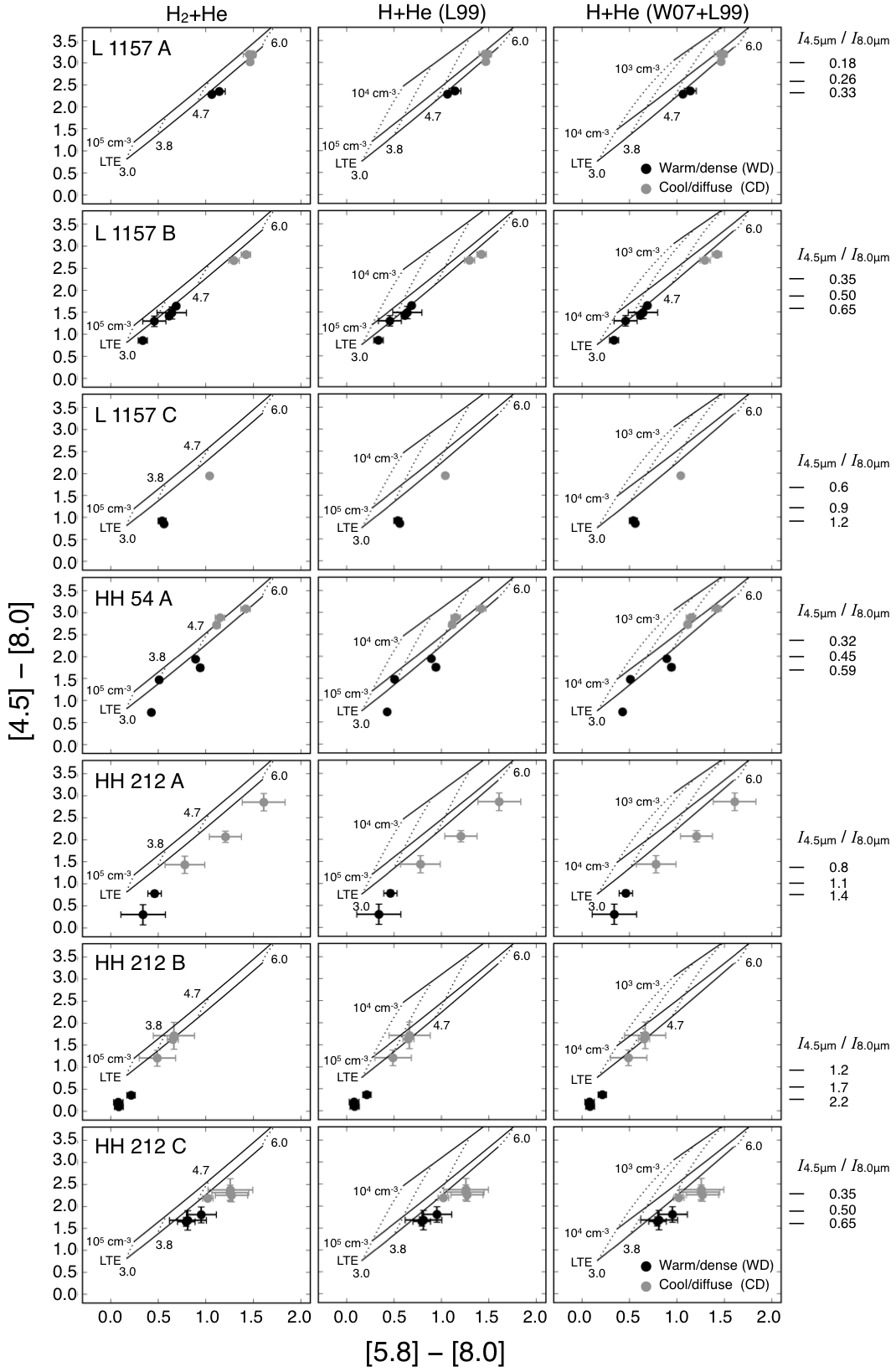}
\end{figure*}

\clearpage

\figcaption{Same as Figure \ref{fig10}, but shock models with a power-law cooling function ($\Lambda \propto T^{-\alpha}$) are plotted instead of using the isothermal assumption. The numbers shown in the plots (3.0/3.8/4.7/6.0) are the power index $\alpha$.
\label{fig11}}

\clearpage

\begin{figure*}
\epsscale{1.9}
\plotone{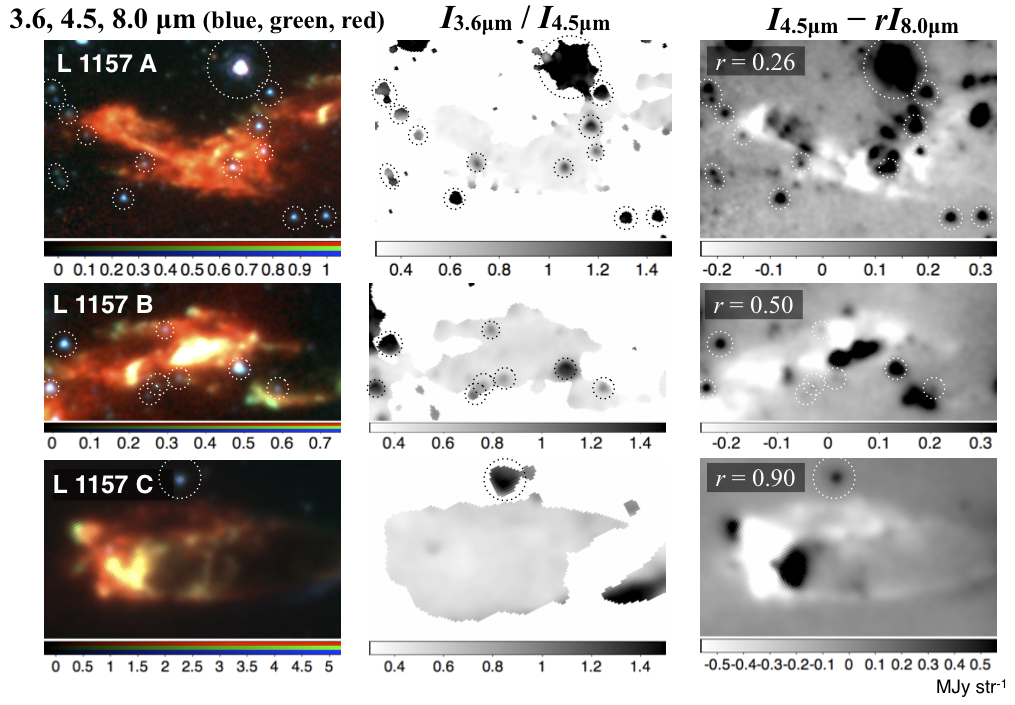}
\caption{
The three-color image, the $I_{3.6 \micron} / I_{4.5 \micron}$ and the $I_{4.5 \micron} - r I_{8.0 \micron}$ maps
for L 1157 A-C. The $I_{3.6 \micron} / I_{4.5 \micron}$ ratio is shown in regions where the signal-to-noise is larger than 10 for $I_{4.5 \micron}$.
\label{figA1}}
\end{figure*}

\clearpage

\begin{figure*}
\epsscale{2.0}
\plotone{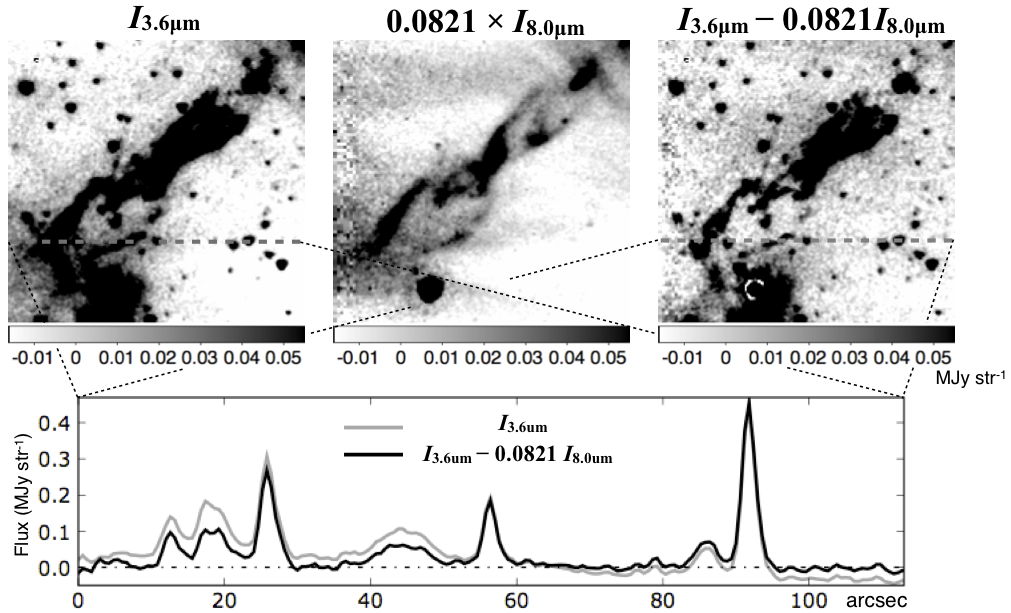}
\caption{
Subtraction of the non-uniform diffuse extended emission at 3.6 \micron~ for part of the HH 212 jet. (top-left) the 3.6-\micron~ image before subtraction. (top-middle) flux-scaled image at 8.0 \micron. (top-right) the 3.6-\micron~ image subtracted from the flux-scaled image at 8.0 \micron. The three figures are displayed with the same contrast level as shown by the scale bars. (bottom) one-dimensional flux distribution extracted from the images before and after subtraction (gray and solid lines, respectively). The dot-dashed line indicates the zero flux level.  
\label{figA2}}
\end{figure*}

\clearpage

\begin{figure*}
\epsscale{1.9}
\plotone{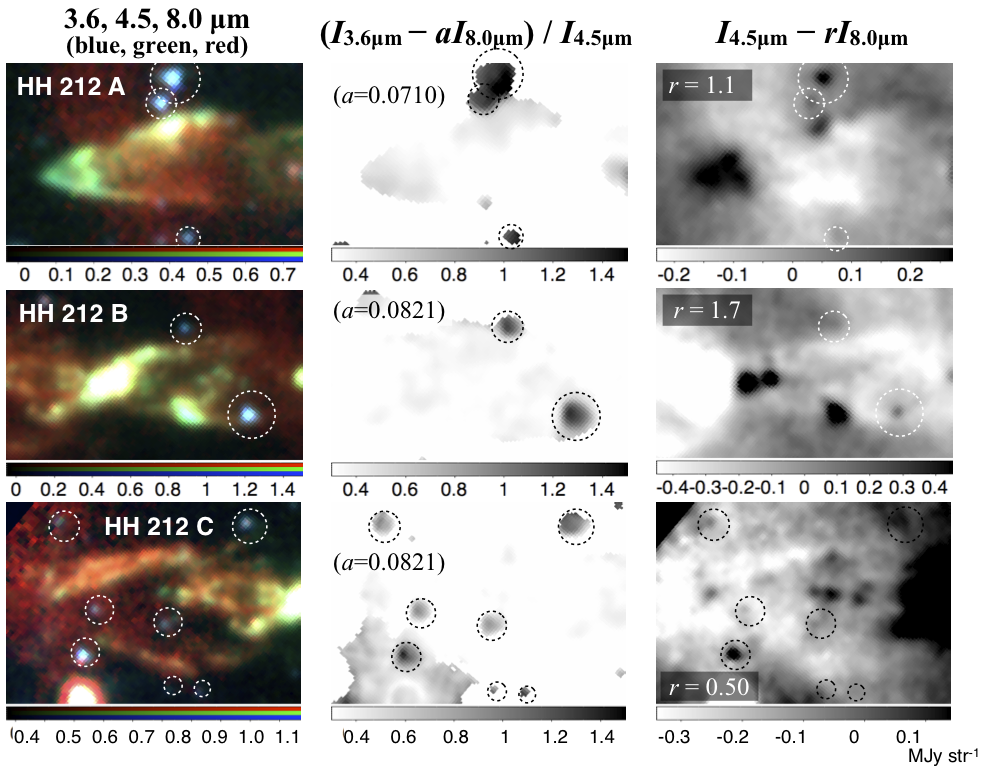}
\caption{
Same as Figure \ref{figA1} but for HH 212 A-C. Non-uniform diffuse extended emission at 3.6 \micron~is subtracted using the flux-scaled image at 8.0 \micron.
\label{figA3}}
\end{figure*}

\begin{deluxetable}{llcccc}
\tablecolumns{6}
\tablecaption{Peak Position of Warm/Dense Knots \label{tbl1}}
\tablehead{
\colhead{Object} & \colhead{Feature} & \colhead{R.A. (2000)}  & \colhead{Dec. (2000)} & \colhead{Distance to the} & \colhead{P.A. from the}\\
\colhead{}             & \colhead{}              &  \colhead{}                      & \colhead{}                       & \colhead{protostar (arcsec)\tablenotemark{a}} & \colhead{protostar (deg.)\tablenotemark{a}}

}
\startdata
L 1157 &  A1\tablenotemark{b}   & 20:38:54.8 & +68:04:57 & 173 & 339 \\
              &         & 20:38:55.9 & +68:04:52 & 166 & 340 \\
              &         & 20:38:56.2 & +68:04:55 & 168 & 341 \\
              &         & 20:38:57.8 & +68:04:48 & 159 & 343 \\
              &  A2\tablenotemark{b}   & 20:39:00.1 & +68:04:21 & 130 & 345 \\
              &         & 20:39:00.8 & +68:04:14 & 122 & 346 \\
              &  A3\tablenotemark{b}   & 20:38:58.9 & +68:04:13 & 124 & 341 \\
              &         & 20:38:58.4 & +68:04:06 & 118 & 338 \\
              &         & 20:38:58.4 & +68:03:58 & 111 & 337 \\
              
              &  B1 (A4)  & 20:38:57.6&+68:03:32 & 90 & 328 \\
              &  B2 (A5) & 20:39:00.5&+68:03:34 & 84 & 338 \\
              &  B3  & 20:38:58.2&+68:03:08 & 68 & 319 \\
              &  B4\tablenotemark{b}   & 20:39:00.6&+68:03:17 & 68 & 333 \\
              &         & 20:39:00.5&+68:03:08 & 61 & 329 \\
              &  B5  & 20:39:04.9&+68:02:55 & 39 & 350 \\

              & C1 & 20:39:09.3 & +68:01:20 & 59 & 163 \\
              & C2 & 20:39:12.1 & +68:01:11 & 73& 153\vspace{0.1cm} \\


HH 212 &  A1\tablenotemark{b}   & 05:43:53.9 & --01:01:29 & 92 & 24 \\
               &         & 05:43:53.7 & --01:01:32 & 88 & 23 \\
               &         & 05:43:53.8 & --01:01:35 & 86 & 25 \\
               &  A2  & 05:43:52.8 & --01:01:44 & 72 & 17 \\

               &  B1\tablenotemark{b}   & 05:43:49.0 & --01:04:09 & 83 & 205\\
               &         & 05:43:49.2 & --01:04:04 & 78 & 205\\
               &  B2  & 05:43:49.2 & --01:03:49 & 65 & 211\\

               &  C1  & 05:43:47.7 & --01:04:58 & 137 & 204 \\
               &  C2  & 05:43:48.9 & --01:04:36 & 109 & 200 \\
               &  C3\tablenotemark{b}  & 05:43:48.3 & --01:04:36 & 113 & 204 \\
               &         & 05:43:48.4 & --01:04:33 & 109 & 204 \\
               &         & 05:43:48.5 & --01:04:27 & 103 & 205 \\
               &         & 05:43:48.7 & --01:04:26 & 101 & 203\vspace{0.1cm} \\


HH 54  & A1 & 12:55:54.9 & --76:56:06 & --- & --- \\
              & A2 & 12:55:51.0 & --76:56:21 & --- & --- \\
              & A3 & 12:55:53.7 & --76:56:22 & --- & --- \\
              & A4 & 12:55:49.5 & --76:56:30 & --- & --- \\

\enddata
\tablenotetext{a}{We adopt ($\alpha_{J2000}$,$\delta_{J2000}$)=(20:39:06.2,68:02:16) and (05:43:51.4,--01:02:53) for the L 1157 jet and HH 212 driving sources, respectively \citep{Bachiller01,LeeC08}. The driving source of HH 54 is not known
(Caratti o Garatti et al. 2006; Paper I)
}
\tablenotetext{b}{These features contains multiple peaks.}
\end{deluxetable}

\clearpage

\begin{deluxetable}{llcccc}
\tablecolumns{6}
\tablecaption{Fluxes measured in warm/dense and cool/diffuse regions \label{tbl2}}
\tablehead{
\colhead{Region} & \colhead{Position} & \multicolumn{4}{c}{Flux (MJy str$^{-1}$)}  \\
\colhead{} & \colhead{} & \colhead{3.6 \micron} & \colhead{4.5 \micron} & \colhead{5.8 \micron} & \colhead{8.0 \micron}
}
\startdata
L 1157 A & WD-A1	& 0.115 (0.005)		& 0.304 (0.006) 	& 0.59 (0.03) 		& 0.95 (0.03) \\                   
		& WD-A2	& 0.216 (0.005) 	& 0.488 (0.005) 	& 0.962 (0.009)		& 1.43 (0.02) \\                   
		& CD-A1	& 0.054 (0.005) 	& 0.178 (0.006) 	& 0.56 (0.03) 		& 1.20 (0.03) \\
		& CD-A2a	& 0.106 (0.005) 	& 0.298 (0.005) 	& 0.802 (0.009)		& 1.72 (0.02) \\                    
		& CD-A2b	& 0.047 (0.005) 	& 0.138 (0.005) 	& 0.422 (0.009)		& 0.93 (0.02)\vspace{0.2cm} \\                   

L 1157 B 	& WD-B1	& 0.097 (0.011) 	& 0.186 (0.007) 	& 0.259 (0.012)		& 0.22 (0.02) \\
		& WD-B2 	& 0.386 (0.011) 	& 0.751 (0.007)  	& 1.163 (0.012)		& 1.22 (0.02) \\
		& WD-B3	& 0.112 (0.008) 	& 0.207 (0.010) 	& 0.29 (0.02) 		& 0.29 (0.03) \\                  
		& WD-B4	& 0.847 (0.008) 	& 1.651 (0.010) 	& 2.21 (0.02) 		& 2.17 (0.03) \\
		& WD-B5	& 0.213 (0.004) 	& 0.474  (0.007) 	& 0.492 (0.010)		& 0.37 (0.02) \\
		& CD-B2	& 0.077 (0.011) 	& 0.154 (0.007)  	& 0.353 (0.012)		& 0.73 (0.02) \\                  
		& CD-B3B4& 0.114 (0.008) 	& 0.258 (0.010) 	& 0.59 (0.02) 		& 1.08 (0.03)\vspace{0.2cm} \\                   

L 1157 C 	& WD-C1     & 2.383 (0.009) 	& 5.766 (0.013) 	& 4.85 (0.03) 	& 4.53 (0.05) \\                   
		& WD-C2     & 0.775 (0.009) 	& 1.768 (0.013)  	& 1.61 (0.03) 	& 1.47 (0.05) \\
		& CD             & 0.668 (0.009) 	& 1.162 (0.013) 	& 1.72 (0.03) 	& 2.49 (0.05)\vspace{0.2cm} \\                   

HH 54 A  	& WD-A1     	& 0.89 (0.04) 	& 2.26 (0.03) 	& 3.05 (0.09) 	& 4.04 (0.04) \\
		& WD-A2     	& 1.21 (0.04)  	& 3.17 (0.03)  	& 4.94 (0.09)  	& 4.39 (0.04) \\
		& WD-A3     	& 2.06 (0.04)   	& 4.05 (0.03)  	& 6.83 (0.09)  	& 8.66 (0.04) \\
		& WD-A4     	& 1.53 (0.04)   	& 4.52 (0.03)  	& 3.84 (0.09)  	& 3.17 (0.04) \\
		& CD-1     		& 0.23 (0.04) 	& 0.63 (0.03)	& 1.99 (0.09)  	& 3.19 (0.04) \\
		& CD-2     		& 0.37 (0.04) 	& 0.99 (0.03)	& 2.79 (0.09)  	& 4.33 (0.04) \\
		& CD-3     		& 0.28 (0.04) 	& 0.64 (0.03)	& 1.91 (0.09)  	& 3.94 (0.04)\vspace{0.2cm} \\
                 
HH 212 A	& WD-A1		& 0.25 	(0.04) 	& 0.52 	(0.03) 	& 0.32	(0.03)	& 0.25	(0.05) \\
		& WD-A2		& 0.434 	(0.010) 	& 0.87	(0.02)	& 0.75	(0.02)	& 0.64	(0.04) \\
		& CD-1		& 0.06 	(0.02)	& 0.168	(0.013)	& 0.24	(0.03) 	& 0.40	(0.04) \\
		& CD-2		& 0.02 	(0.02)	& 0.081	(0.013)	& 0.16	(0.03) 	& 0.40	(0.04) \\
		& CD-A2		& 0.17 	(0.02)   	& 0.39	(0.04)	& 0.46	(0.05)	& 0.52	(0.08)\vspace{0.2cm} \\

HH 212 B	& WD-B1a	& 1.43 	(0.03)   	& 4.00    	(0.06)	& 2.95    	(0.07)	& 2.00	(0.07) \\
		& WD-B1b	& 1.48 	(0.03)    	& 3.96    	(0.06)	& 2.85    	(0.07)	& 1.70 	(0.07) \\
		& WD-B2		& 0.738 	(0.008)  	& 2.02    	(0.02)	& 1.31    	(0.04)	& 0.79	(0.03) \\
		& CD-1		& 0.08 	(0.02)	& 0.18	(0.04)	& 0.30	(0.04)	& 0.31	(0.05) \\
		& CD-2		& 0.176 	(0.009)   	& 0.388 	(0.015)	& 0.48	(0.04)	& 0.42	(0.07) \\
		& CD-B2		& 0.108 	(0.008)   	& 0.267	(0.013)	& 0.42	(0.02)	& 0.43	(0.02)\vspace{0.2cm} \\

HH 212 C	& WD-C2		& 0.153 	(0.012)  	& 0.306	(0.02)	& 0.43	(0.03)	& 0.50	(0.03) \\
		& WD-C3a	& 0.12	(0.05)	& 0.347	(0.04)	& 0.49	(0.05)	& 0.66	(0.07) \\
		& WD-C3b	& 0.10 	(0.05)	& 0.302	(0.04)	& 0.43	(0.05)	& 0.51	(0.07) \\
	 	& CD-1		& 0.085	(0.006)	& 0.233	(0.011)	& 0.44	(0.02)	& 0.62	(0.02) \\
		& CD-2		& 0.059	(0.015)	& 0.13	(0.02)	& 0.23	(0.03)	& 0.40	(0.07) \\
		& CD-C3a		& 0.079	(0.013)	& 0.224	(0.016)	& 0.36	(0.04)	& 0.64	(0.07) \\
		& CD-C3b		& 0.063	(0.013)	& 0.184	(0.016) 	& 0.31	(0.04)	& 0.55	(0.07)
\enddata
\end{deluxetable}


%

\begin{deluxetable}{lrrllll}
\tablecolumns{7}
\rotate
\tabletypesize{\scriptsize}
\tablecaption{Comparison of IRAC Data with IRS Observations and IRAC Model Emission for the L 1157 Protostellar Jet \label{tbl3}}
\tablehead{
\colhead{Position}  & \multicolumn{2}{c}{LTE temperature (K)} & \multicolumn{2}{c}{IRAC 8.0-\micron~flux } &  \multicolumn{2}{c}{IRAC 4.5-\micron~flux} \\

\colhead{}                & \multicolumn{1}{c}{IRS} & \multicolumn{1}{c}{IRAC}  & \multicolumn{1}{c}{modeled\tablenotemark{b}}  & \multicolumn{1}{c}{modeled/observed\tablenotemark{c}} & \multicolumn{1}{c}{modeled\tablenotemark{d}}  & \multicolumn{1}{c}{modeled/observed\tablenotemark{e}}\\ 
                
\colhead{}                &\multicolumn{1}{c}{S(5)--S(7)  lines\tablenotemark{a}} & \multicolumn{1}{c}{$I_{5.8\micron}/ I_{8.0\micron}$}	& \multicolumn{1}{c}{(MJy str$^{-1}$)} &  & \multicolumn{1}{c}{(MJy str$^{-1}$)} &
}
\startdata
WD-A2	& 1200		&	1070$\pm$10  	&  $1.2 \pm 0.3$	& $0.8 \pm 0.5$	&  $0.14 \pm 0.04$ 	 	&  $0.29 \pm 0.11$\\
CD-A2a	&   900		&	920$\pm$20 	&  $1.4 \pm 0.4$	& $0.8 \pm 0.5$	&  $0.033 \pm 0.008$	&  $0.11 \pm 0.05$\\
CD-A2b	&   860		&	910$\pm$20 	&  $0.5 \pm 0.2$	& $0.5 \pm 0.4$     	&  $0.008 \pm 0.003$	&  $0.06 \pm 0.03$\vspace{0.1cm}\\

WD-B2	& 1060		&	1250$\pm$20   &  $0.9 \pm 0.3$	& $0.7 \pm 0.5$	&  $0.08 \pm 0.02$	& $0.11 \pm 0.05$\\
CD-B2	& ---			& 	940$\pm$20    	&  $0.7 \pm 0.3$	& $1.0 \pm 0.8$	&  ---		& ---\vspace{0.1cm}\\

WD-B4	& 1300		&	1290$\pm$20   &  $2.2 \pm 0.6$	& $1.0 \pm 0.7$   	&  $0.5 \pm 0.1$			& $0.30 \pm 0.12$\\
CD-B3B4	&   890		&	980$\pm$30 	&  $0.9 \pm 0.3$	& $0.8 \pm 0.6$	&  $0.021 \pm 0.005$	& $0.08 \pm 0.04$\vspace{0.1cm}\\

WD-C1	& 1350		&	1330$\pm$10   &  $4.9 \pm 1.2$	& $1.1 \pm 0.7$	&  $1.2 \pm 0.3$		& $0.21 \pm 0.09$\\
WD-C2	& 1500		&	1340$\pm$40   &  $0.7 \pm 0.2$	& $0.5 \pm 0.3$	&  $0.26 \pm 0.06$		& $0.15 \pm 0.05$\\
CD-C1C2	& 1100		&	1080$\pm$20	&  $2.4 \pm 0.6$	& $1.0 \pm 0.5$	&  $0.20 \pm 0.05$		& $0.17 \pm 0.08$\\
\enddata
\tablenotetext{a}{Based on Figure 5 in \citet{Nisini10}, with an uncertainty of $\sim$50 K.}
\tablenotetext{b}{Based on Equation 2 of \citet{Neufeld08}, assuming that the flux is dominated by H$_2$ 0--0 S(4) and S(5) lines. The uncertainty includes systematic error of the observed IRS fluxes \citep[up to 25 \%,][]{Neufeld06}.}
\tablenotetext{c}{The systematic error associated with photometric calibration is assumed to be within 40 \% of the observed flux based on the IRAC Data Handbook 3.0 (see text).}
\tablenotetext{d}{assuming that (1) the flux is dominated by the H$_2$ 0--0 S(9) line; and (2) LTE at the temperature measured using the 0--0 S(5), S(6) and S(7) lines. The uncertainty includes a systematic error of the observed IRS fluxes of  25 \%.}
\tablenotetext{e}{The systematic error associated with photometric calibration is assumed to be within 20 \% of the observed flux based on the IRAC Data Handbook 3.0 (see text).}
\end{deluxetable}




\end{document}